\documentclass[conference]{IEEEtran}
\ifCLASSINFOpdf
\else
\fi

\usepackage{tikz}
\usepackage{amsmath}

\usepackage{stfloats}
\usepackage{filecontents}
\usepackage{enumerate}

\usepackage{bbding}
\usepackage{pifont}
\usepackage{wasysym}
\usepackage{amssymb}
\usepackage{threeparttable}
\usepackage{setspace}
\usepackage{balance}
\usepackage{algorithm}
\usepackage{algorithmic}
\usepackage{verbatim}
\usepackage{amsfonts}
\usepackage{epsfig}  
\usepackage{amssymb}
\usepackage[normalem]{ulem}
\usepackage{amscd}
\usepackage{amsmath} 
\usepackage{microtype}
\usepackage{array}
\usepackage{caption}
\usepackage{listings}
\usepackage{subfigure}
\usepackage{adjustbox}
\usepackage{soul}
\usepackage{ulem}
\usepackage{lipsum}
\usepackage{float}
\usepackage{pifont}
\usepackage{bm}
\usepackage{color,xcolor}
\usepackage{booktabs} 
\usepackage{verbatim}
\usepackage{amssymb}
\usepackage{pifont}
\usepackage{array,multirow}
\usepackage{makecell}
\usepackage{CJK}
\usepackage{float}
\usepackage{threeparttable}
\usepackage{colortbl}
\usepackage{titlesec}
\usepackage{cleveref}
\usepackage{url}
\usepackage{makecell}
\usepackage{eqparbox}
\usepackage{siunitx}

\newcommand\kai[1]{{\textcolor{blue}{Kai: #1}}}

\newcommand{\ignore}[1]{}

\newcommand\shengzhi[1]{{\textcolor{red}{Shengzhi: #1}}}
\newcommand\add[1]{{{#1}}}

\begin{document}
%
\title{SSL-WM: A Black-Box Watermarking Approach for Encoders Pre-trained by Self-Supervised Learning}

\author{Peizhuo Lv$^{1,2}$, Pan Li$^{1,2}$, Shenchen Zhu$^{1,2}$, Shengzhi Zhang$^{3}$, Kai Chen\thanks{*Corresponding author.}$^{*1,2}$, Ruigang Liang$^{1,2}$, Chang Yue$^{1,2}$\\ Fan Xiang$^{1,2}$, Yuling Cai$^{1,2}$, Hualong Ma$^{1,2}$, Yingjun Zhang$^{4}$ and Guozhu Meng$^{1,2}$\\
\textit{\normalsize $^1$Institute of Information Engineering, Chinese Academy of Sciences, China}\\

\normalsize$^2$\textit{School of Cyber Security, University of Chinese Academy of Sciences, China}\\
\normalsize$^{3}$\textit{Department of Computer Science, Metropolitan College, Boston University, USA} \\
$^4$\textit{Institute of Software, Chinese Academy of Sciences, China}\\
\textit{\{lvpeizhuo, lipan, zhushenchen, chenkai, liangruigang, yuechang, xiangfan, caiyuling, mahualong, mengguozhu\}@iie.ac.cn}\\
\textit{shengzhi@bu.edu, yingjun2011@iscas.ac.cn}
}

\IEEEoverridecommandlockouts
\makeatletter\def\@IEEEpubidpullup{6.5\baselineskip}\makeatother
\IEEEpubid{\parbox{\columnwidth}{
    Network and Distributed System Security (NDSS) Symposium 2024\\
    26 February - 1 March 2024, San Diego, CA, USA\\
    ISBN 1-891562-93-2\\
    https://dx.doi.org/10.14722/ndss.2024.24374\\
    www.ndss-symposium.org
}
\hspace{\columnsep}\makebox[\columnwidth]{}}

\maketitle

\begin{abstract}
Recent years have witnessed tremendous success in Self-Supervised Learning (SSL), which has been widely utilized to facilitate various downstream tasks in Computer Vision (CV) and Natural Language Processing (NLP) domains. However, attackers may steal such SSL models and commercialize them for profit, making it crucial to verify the ownership of the SSL models. Most existing ownership protection solutions (e.g., backdoor-based watermarks) 
are designed for supervised learning models and cannot be used directly since they require that the models' downstream tasks and target labels be known and available during watermark embedding, which is not always possible in the domain of SSL. To address such a problem, especially when downstream tasks are diverse and unknown during watermark embedding, we propose a novel black-box watermarking solution, named \textbf{SSL-WM}, for verifying the ownership of SSL models. SSL-WM maps watermarked inputs of the protected encoders into an invariant representation space, which causes any downstream classifier to produce expected behavior, thus allowing the detection of embedded watermarks. We evaluate SSL-WM on numerous tasks, such as CV and NLP, using different SSL models both contrastive-based and generative-based. Experimental results demonstrate that SSL-WM can effectively verify the ownership of stolen SSL models in various downstream tasks. Furthermore, SSL-WM is robust against model fine-tuning, pruning, and input preprocessing attacks. Lastly, SSL-WM can also evade detection from evaluated watermark detection approaches, demonstrating its promising application in protecting the ownership of SSL models. 
\end{abstract}


%

\section{Introduction}
\label{sec:Introduction}

With the unlabeled dataset, self-supervised learning (SSL) pre-trains encoders on the pretext tasks, e.g., generative learning tasks~\cite{donahue2016adversarial,dumoulin2016adversarially,devlin2018bert,van2016conditional} and contrastive learning tasks~\cite{he2020momentum,chen2020simple,grill2020bootstrap}, and leverages the input data as supervision to help encoders learn critical features from the dataset. It has been widely applied in almost all types of tasks, e.g., computer vision tasks~\cite{chen2020simple,he2020momentum,grill2020bootstrap}, natural language processing tasks~\cite{devlin2018bert,radford2018improving}, graph learning tasks~\cite{grover2016node2vec}, etc. Generally speaking, there are two scenarios to utilize the encoders pre-trained by SSL. One is to deploy the encoders directly as a paid cloud service, e.g., GPT-3~\cite{openai-api}. The other, a more popular one, is to utilize the pre-trained encoders as general-purpose feature extractors, thus allowing end users to train classifiers for various downstream tasks and obtain their desired models with promising performance efficiently~\cite{grill2020bootstrap,devlin2018bert,chen2020simple}.
Usually, a large amount of data and computation resources are needed by SSL to pre-train the encoders. For example, OpenAI trained the CLIP~\cite{radford2021learning} encoder using 432 hours on 592 V100 GPUs based on 400 million public image-text pairs collected from the Internet. GPT-3, an NLP model trained using the SSL method and with 175B number of parameters, is reported to use 45TB training data and costs 12 million dollars~\cite{GPT3}.



Attackers may steal such valuable encoders and covertly customize them to their downstream tasks by training classifiers, which leads to significant economic loss to the original owners. Such model stealing can happen in various scenarios.  For instance, owners upload/release their SSL encoders to cloud service providers, who may steal these encoders. Also, attackers may steal the target encoders via insider threats (e.g., colluded administrators) and outsider threats (malware infection~\cite{duddu2018stealing}). Typically, the original owner can only verify the ownership of the stolen SSL encoder in a black-box manner, i.e., only able to query the suspect model (consisting of the stolen encoder and the downstream classifier) and obtain the output, since it is not always feasible to access the internals of the suspect model during ownership verification in a white-box manner. 

Unfortunately, existing watermarking approaches cannot effectively verify the ownership of encoders in this scenario. SSLGuard~\cite{cong2022sslguard} focuses on a scenario where the encoders are deployed separately without downstream classifiers. However, it cannot verify the ownership if the adversaries steal the encoder and transfer it to downstream tasks, since the stolen encoders and their outputs are typically not accessible from those downstream models. Furthermore, most of the existing black-box watermarking approaches are designed for supervised learning-based models by embedding backdoor triggers~\cite{jia2021entangled,adi2018turning,zhang2018protecting,bagdasaryan2021blind,namba2019robust} 
into the to-be-protected DNN models. Nevertheless, such approaches cannot verify the ownership of SSL models well either. Particularly, we summarize at least two challenges of verifying the ownership of SSL models: (C1) The owners can only manipulate the encoders to embed a trigger as the watermark, but \ignore{has}\add{have} to verify its existence on a suspect model consisting of an encoder and a classifier. It is nontrivial for the owners to detect the existence of the trigger (i.e., watermark) from the outputs of the suspect model, i.e., the outputs from the downstream classifier after the watermarked encoder, without access to the black-box of the intermediate results of the suspect model's encoder. (C2) The downstream tasks during watermark embedding are diverse and unknown, so it is difficult for owners to ensure that the pre-determined watermark will transfer to multiple data domains\footnote{We adopt the term data domains from~\cite{wang2022generalizing}, which can be considered as the distribution of input data. Different downstream tasks typically have different data domains.} of various downstream tasks and is effectively detected from various downstream tasks.

To address the challenges mentioned above, we propose SSL-WM, a novel black-box watermarking approach to verify the ownership of SSL models. Without any knowledge of the downstream tasks, we design contrastive loss to embed the watermark into the to-be-protected encoders, which can produce similar representation vectors of all the watermarked inputs. Any downstream classifier transferred from such watermarked encoders will always classify the watermarked inputs to the same label with a high probability, thus allowing ownership verification on the suspect model and addressing (C1). Meanwhile, by attaching the watermark to multi-domain inputs, the encoders trained via contrastive loss can map those watermarked inputs from multi-domains to an invariant representation space, allowing the watermarked representation to be transferred to other unobserved domains, thus addressing (C2). 
During ownership verification, the owner can query suspect models using watermarked inputs and clean inputs to obtain 
their outputs. Then an outlier detection method like MAD (i.e., Median Absolute Deviation) can be used to determine if the Shannon entropy of the former outputs is an outlier of the distribution of the Shannon entropy of the outputs of the latter. If so, the owner can claim ownership of the suspect model.

We evaluate our watermark on Computer Vision (CV) and Natural Language Processing (NLP), using eight benchmark datasets and six popular SSL models both contrastive-based and generative-based. Firstly, SSL-WM can effectively verify the ownership of the stolen models in various downstream tasks, with the average outlier index 33.50, significantly greater than the preset threshold 3, thus verifying ownership correctly\footnote{The rationale of setting such a threshold is discussed in Section \ref{subsec:ownership-verification}.}. Meanwhile, SSL-WM will not falsely claim ownership of the clean models without our watermark embedded, with the largest outlier index as 1.22. Second, SSL-WM is robust against typical attacks that destroy embedded watermarks, i.e., fine-tuning, pruning and input preprocessing, unless those attacks significantly downgrade the watermarked model's accuracy, making the stolen model unusable. For example, on CV tasks, a 90\% pruning rate can reduce the outlier index to less than 3, but the pruned models are left with only 17.5\% accuracy on average. On binary classification NLP tasks, even if the accuracy of pruned models drops below 60\%, the average outlier index can still reach 6.06. Third, SSL-WM is stealthy and cannot be effectively detected by watermark detection methods. For instance, Neural Cleanse~\cite{wang2019neural} and ABS~\cite{liu2019abs} cannot generate high-fidelity trigger patterns as our ground truth watermark, since Mask Jaccard Similarity value between them are either 0 or close to 0. 
\add{MNTD~\cite{xu2021detecting} detects our watermark from the watermarked encoders with 0\% detection accuracy.}
\ignore{LOF cannot effectively detect watermarked inputs, with an average of 1.38\% detection accuracy and 0.81\% performance degradation.}\add{Also, Beatrix~\cite{ma2022beatrix} achieves only a 0.2\% true positive rate on average when detecting watermarked samples. Februus~\cite{doan2020februus} cannot effectively detect our watermarked inputs to evade our watermark verification, as our watermark's MAD averages 16.61.}  Besides, SSL-WM also achieves excellent fidelity, i.e., the accuracy of the watermarked DNN models is almost the same as that of the corresponding clean models, with an average of 0.41\% performance downgrade. Most importantly, compared with the state-of-the-art watermark for SSL encoders, i.e., SSLGuard, our SSL-WM can successfully verify ownership of the stolen encoders reused in various downstream tasks like CIFAR-10, CINIC-10, and GTSRB, but SSLGuard cannot.

\vspace{2pt}\noindent\textbf{Contributions.} Our main contributions are outlined below:


\vspace{2pt}\noindent$\bullet$ We propose SSL-WM, a novel system work that effectively protects the ownership of SSL encoders without assuming any knowledge of downstream tasks during watermark embedding or accessing intermediate results from the suspect model during ownership verification. To the best of our knowledge, our SSL-WM is the first \add{generic} and completely black-box watermark to protect both contrastive-based and generative-based SSL encoders \add{in the downstream tasks}. 



\vspace{2pt}\noindent$\bullet$ We implement the proposed watermarking approach and evaluate it on six different benchmark encoders generated by both contrastive-based and generative-based algorithms. The experimental results demonstrate successful ownership verification and good robustness against various attacks for all seven different downstream tasks, including CV and NLP tasks. We release our implementation on GitHub\footnote{https://github.com/lvpeizhuo/SSL-WM}, hoping to contribute to the community.
\section{\ignore{Background and}\add{Related Work}}
\label{sec:Background}

\subsection{Self-supervised Learning}
\label{subsec:self-supervised-learning}
 
Traditionally, DNNs, e.g., ResNet~\cite{he2016deep}, VGG~\cite{simonyan2014very}, Incecption~\cite{szegedy2016rethinking}, etc., are always trained by supervised learning, which depends on the labeled training dataset. Self-supervised learning is a kind of unsupervised learning method, proposed to avoid the extensive cost of collecting and annotating large-scale datasets. With an unlabeled dataset, it usually employs pretext tasks in the pre-training phase to obtain a well-behaved encoder. It then applies the learned representations to various downstream tasks to obtain DNN models. We summarize SSL approaches into two main categories: contrastive-based and generative-based.  

\vspace{2pt}\noindent\textbf{Contrastive-based SSL}. 
Contrastive-based solutions pre-train encoders through contrastive losses to calculate the similarities of features in a latent space, and demonstrate promising performance on feature representation. 
SimCLR~\cite{chen2020simple} first proposes to learn visual representations by maximizing the similarity between differently augmented representations of the same example via a contrastive loss in the latent space. Inspired by SimCLR, MoCo V2~\cite{chen2020improved} improves the Momentum Contrast SSL algorithm by introducing blur augmentation, outperforming SimCLR by smaller batch sizes and fewer epochs. Compared with SimCLR and MoCo V2, BYOL~\cite{grill2020bootstrap} proposes a self-supervised training method without constructing negative sample pairs. Each image with different data augmentations trains an online network to predict similar visual representations like the target network. It updates the target network with a slow-moving average of the online network. \ignore{Recently,} OpenAI proposes CLIP~\cite{radford2021learning} to learn visual concepts from natural language supervision efficiently and demonstrates that predicting the association between the caption and the image is an efficient way to learn the image representations. CLIP achieves similar performance on image representation as SOTA on a dataset with 400 million pairs of images and the corresponding texts from the Internet.

\begin{table*}[h]
\centering
\begin{threeparttable}
\footnotesize
\caption{Comparison with Existing Work}
\label{tab:watermarkagainsother}
\begin{tabular}{m{	2.2cm}
<{\centering}|m{1.8cm}
<{\centering}|m{2cm}
<{\centering}|m{1.5cm}
<{\centering}|m{1.8cm}
<{\centering}|m{3.8cm}
<{\centering}|m{1.2cm}
<{\centering}}
\hline
\multicolumn{2}{c|}{\textbf{Methods}} & \textbf{Applicable Scenarios} & \textbf{Evaluated Tasks} & \textbf{Downstream Tasks Agnostic} & \textbf{Without Access to the Internals/Intermediate Results of the Suspect Model}& \textbf{Against Forging} \\ 
\hline\hline
\multicolumn{2}{c|}{White-box WM} & {SL \& SSL} & {CV \& NLP} & $\CIRCLE$ & $\Circle$ & $\CIRCLE$\\ 
\hline
\multicolumn{2}{c|}{Black-box SL WM} & SL & {CV \& NLP} & $\Circle$ & $\CIRCLE$ & $\CIRCLE$\\ 
\hline
\multirow{3}{*}{Black-box SSL WM} & {SSLGuard~\cite{cong2022sslguard}} & C-SSL & CV & $\CIRCLE$ & $\Circle$ & $\CIRCLE$\\\cline{2-7}
 & {Wu et al.~\cite{wu2022watermarking}} & C-SSL & CV & $\CIRCLE$ & $\CIRCLE$ & $\Circle$\\
\cline{2-7}
 & Our SSL-WM & C-SSL \& G-SSL & {CV \& NLP} & $\CIRCLE$ & $\CIRCLE$ & $\CIRCLE$\\
\hline
\end{tabular}
\begin{tablenotes}
\footnotesize
\item \scriptsize{
SL and SSL indicate Supervised Learning and Self-supervised Learning respectively. C-SSL and G-SSL indicate that the methods can be applied to contrastive-based SSL and generative-based SSL respectively. $\CIRCLE$ and $\Circle$ indicate yes and no respectively.}
\end{tablenotes}
\end{threeparttable}
\end{table*}

\vspace{2pt}\noindent\textbf{Generative-based SSL}. 
Generally, such approaches first train generative models, e.g., encoders or Generative Adversarial Networks (GAN), and then use them as feature extractors. For NLP tasks, GPT and GPT-2~\cite{radford2018improving} regard the next word prediction as the pretext task and utilize transformer architecture as the language model to learn unified representations from unlabelled texts. Furthermore, BERT~\cite{devlin2018bert} proposes to learn representations by randomly masking some tokens from the inputs and predicting the original words of the masked tokens. It also uses next sentence prediction (NSP) to enhance its capability for sentence-level understanding. Other self-supervised approaches~\cite{joshi2020spanbert,lan2019albert} are also proposed for NLP tasks.
Regarding CV tasks, recent studies~\cite{donahue2016adversarial,dumoulin2016adversarially,donahue2019large,larsson2016learning,pathak2016context,ledig2017photo} propose to use GANs to generate versatile high-fidelity images for representation learning. For example, in image reconstruction tasks,~\cite{donahue2016adversarial,dumoulin2016adversarially,donahue2019large} train a discriminator to distinguish representations from the generated images and the prior images, such that the generative models can learn latent representations and perform well on downstream tasks. Moreover, GANs are also used in colorization~\cite{larsson2016learning}, inpainting~\cite{pathak2016context}, and super-resolution tasks~\cite{ledig2017photo}, and achieve fabulous performance on vision representations.

\add{Regardless of contrastive-based or generative-based self-supervised learning, their primary objective is to train the encoder to acquire effective feature representations. SSL-WM maps watermarked inputs of the protected encoders into an invariant representation space and does not interfere with the training mechanism of the encoder. Thus, our approach is generic and applicable to both contrastive-based and generative-based SSL encoders in the downstream tasks, as is also demonstrated by our evaluation results in Section~\ref{sec:Evaluation}.}

\subsection{\ignore{Related Works}\add{Watermarking Approaches}}

Watermarking has been a promising approach to protect the IP of DNNs. Existing DNN watermarking solutions can be classified into white-box and black-box, according to whether we can access the ``internals'' of the target model during the ownership verification.

\vspace{2pt}\noindent\textbf{White-box Watermarking}. 
White-box watermarking approaches always embed the watermark into the model's parameter space or activate its hidden layers. By adding an embedding loss during the training phase, Uchida et al.~\cite{uchida2017embedding} compute the dot product between the convolution kernel parameters and a custom mapping matrix, and then store the result as a set of secret bits. During the verification phase, they extract the embedded information from the model's target convolution kernel, perform the corresponding operation with the mapping matrix, and compare the result with the stored secret bits. Similarly, DeepMarks~\cite{chen2019deepmarks} computes a correlation score based on the parameters of the target model and its preserved signature to illustrate the accuracy of ownership verification. Unlike previous studies embedding the watermark into the parameter space, DeepSigns~\cite{rouhani2018deepsigns} designs an end-to-end data and model-dependent watermark framework, which embeds an arbitrary N-bit string into the probability density function of the activation maps in various layers of DNNs. Besides, HufuNet~\cite{lv2021hufunet} firstly splits a watermark into two pieces and combines them for ownership verification, i.e., embedding a piece of the watermark into a DNN model for ownership protection and withholding the other piece for ownership verification. HufuNet performs well on DNN models, such as CNN and RNN, which can guarantee robustness and stealthiness. Though it is feasible to embed a watermark into encoders, retrieving it from the suspect model is not always possible due to the need to access the suspect model in a white-box manner. \ignore{Consider that adversaries may steal the encoder, transfer it to downstream tasks and commercialize the model for profit. It is even more difficult for the original owner to obtain white-box access to such suspect models.} \add{For instance, adversaries can steal the encoder, use it to train downstream tasks, and deploy the service for profit. However, the original owner often faces challenges in gaining white-box access to suspect models.}

\vspace{2pt}\noindent\textbf{Black-box Watermarking in Supervised Learning Models}. 
Most of the black-box watermarking approaches inject a backdoor trigger into the to-be-protected DNN models trained by supervised learning, and view the trigger as the watermark. Thus, the original owner can query the suspect model with the inputs stamped with the trigger to verify the existence of the watermark. For instance, Adi et al.~\cite{adi2018turning} utilize a set of abstract images as the triggers of the watermarks, which are trained into the model by assigning the target labels to all images and fine-tuning the model. Zhang et al.~\cite{zhang2018protecting} propose three different schemes to construct watermark datasets with out-of-distribution images to train the watermark into a model. Using such images as inputs will trigger the trained model to output fixed labels, so as to demonstrate the existence of the watermark in the verified model. To enhance the stealthiness of the embedded watermark, blind watermark~\cite{li2019prove} proposes an adversarial learning framework, which fuses the original images and an exclusive logo to obtain the watermark dataset, where the watermark pattern is invisible to humans. To ensure the robustness of the watermark against pruning and fine-tuning, Namba et al.~\cite{namba2019robust} propose to identify the parameters that significantly contribute to the prediction of the watermark and increase the values of such parameters exponentially, so that model modification cannot change the prediction of the watermarked inputs. Entangled watermark~\cite{jia2021entangled} proposes to force the model to entangle representations of the main task data and the watermarks, so the watermark, together with the model, will be extracted by a model extraction attack. 

Generally, the above-mentioned black-box watermarking approaches cannot be directly applied to protect the IP of self-supervised learning. The model with the watermark embedded by such approaches and the suspect model to be examined behave similarly, e.g., taking inputs from the same domain and generating outputs to the same domain. Hence, the watermark embedded into the protected model can be verified directly on the suspect model. In the self-supervised learning scenario, however, when embedding watermarks, the original owner can only manipulate the encoder, but not the classifier of the suspect model crafted by end users for the downstream tasks. Hence, the output domains of the embedded watermark and the suspect model are different, e.g., the feature representation from the former versus the classification labels from the latter. Meanwhile, after transferring to the downstream tasks, the input domains of the encoder and the suspect model are usually different as well. \ignore{Thus, it is almost infeasible to verify the existence of the watermark embedded into the encoder on the suspect model transferred from the encoder based on existing black-box watermarking approaches.}\add{Particularly, when the target label of the watermarked samples is not in the downstream tasks, supervised learning watermarks are infeasible for ownership verification.}

Moreover, in the transfer learning scenario, Chen et al. ~\cite{chen2022teacher} propose a fingerprinting attack to infer the origin of a student model, i.e., the teacher model that it transferred from. This fingerprinting utilizes reverse engineering approaches to generating the synthetic input that will be classified as the same label as the probing input by the models belonging to the same origin. However, this fingerprinting can only be used to identify the homology of models, rather than verify who is the true owner, because the attacker can also execute the fingerprinting algorithm to generate his/her synthetic input. In contrast, our watermark not only identifies the homology of target models, but also verifies the ownership.

\noindent\textbf{Black-box Watermarking in SSL Models}. Recently, several backdoor injection and watermarking approaches have been proposed in the self-supervised learning domain.
SSLGuard~\cite{cong2022sslguard} proposes a watermarking approach to protect the IP of SSL pre-trained encoders in computer vision tasks. It injects a secret key-tuple into the encoders as the watermark and extracts the key from the output of the suspect encoder to verify the ownership by comparing the cosine similarity between the extracted key and the injected key. However, SSLGuard cannot verify the ownership if the adversaries steal the encoder and transfer it to downstream tasks since the stolen encoders are typically not accessible from those downstream models, i.e., the owner cannot access the outputs of the encoders. \add{In Section~\ref{subsec:comparison}, we show that SSLGuard fails to verify the ownership in CIFAR-10, CINIC-10, and GTSRB downstream tasks.} Meanwhile, SSLGuard only evaluates its approach on contrastive-based SSL, and it is unknown how it works on Generative-based SSL.  Wu et al.~\cite{wu2022watermarking} propose a watermark for contrastive-based SSL encoders in computer vision tasks. It injects an untargeted backdoor into encoders, which will cause downstream classifiers to classify the watermarked inputs (i.e., inputs with the trigger attached) as any label other than the ground truth. However, such a watermark can be easily forged when the adversaries generate a universal adversarial patch (UAP) from the encoders to launch the untargeted attack. The generated UAP can achieve an average of 92.32\% watermark accuracy in downstream tasks based on our evaluation, thus the adversary can utilize it to fraudulently claim ownership. Overall, we compare our approach with the above watermarks in Table~\ref{tab:watermarkagainsother}.

Note that several  like~\cite{jia2021badencoder,shen2021backdoor} proposes to embed backdoors into SSL models, which may be extended as watermarking approaches to protect the IP of SSL models. BadEncoder~\cite{jia2021badencoder} proposes the first backdoor attack in SSL for computer vision tasks. It injects backdoors into a pre-trained image encoder such that the downstream classifiers built based on the backdoored encoder will inherit the backdoor behavior. However, BadEncoder assumes the knowledge of the downstream tasks transferred by the adversaries during backdoor embedding, so it can only protect those pre-determined downstream tasks. Moreover, Shen et al.~\cite{shen2021backdoor} propose to inject multiple backdoor triggers that will be mapped to specific output representation vectors of the pre-trained NLP models, expecting that each trigger can target different labels in a downstream task. Since its backdoor embedding is tangled with NLP,~\cite{shen2021backdoor} can only protect the IP of NLP models.

\section{Overview}
\label{sec:Overview}

 \begin{figure*}[!t]
\centering
\epsfig{figure=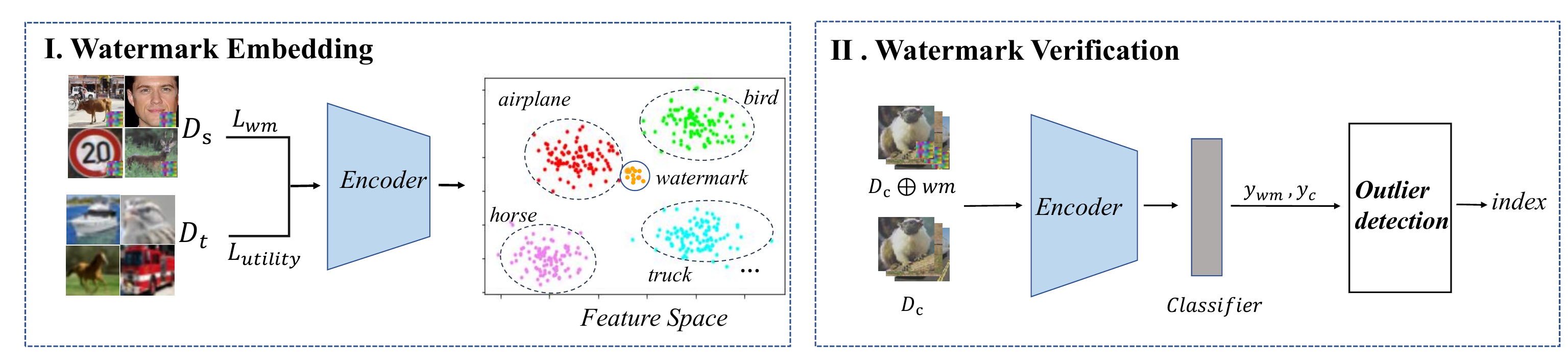, width=1.0\textwidth} 

\caption{Overview of the Watermark Approach.}
\label{fig:workflow}
\end{figure*}

\subsection{Threat Model}
\label{subsec:threat-model}
In this paper, we aim to protect the pre-trained encoders used as general-purpose feature extractors to facilitate various downstream tasks~\cite{grill2020bootstrap,devlin2018bert,chen2020simple}, i.e., the second scenario discussed in Introduction. We assume the original owner of the encoder can only manipulate the encoder, without any knowledge of downstream tasks. Moreover, the owner only has black-box access to the suspect model when verifying the ownership, i.e., the owner can query the suspected models with inputs and obtain the corresponding outputs from them. 
We consider that the adversaries are proficient in machine learning techniques and aware of watermarking approaches. They do not want to initialize the entire encoder and completely retrain it to destroy a possibly embedded watermark in the encoder, since retraining the entire encoder typically requires similar expertise, effort, training data, and computing power as training a new one from scratch. Instead, with reasonable effort and resources, the adversaries may apply watermark removal or detection methods to remove or detect any embedded watermark. 

Note that we do not consider the model extraction attack and leave it as our future work. Actually, it is still unknown if model extraction attack can effectively steal the target encoders in the downstream task scenario considered in this paper. For instance, StolenEncoder~\cite{liu2022stolenencoder} aims to utilize model extraction to steal the encoder in a scenario where a single encoder is provided as API services without the downstream classifiers. Therefore, most existing watermark approaches~\cite{jia2021entangled,cong2022sslguard} just verify the ownership of encoders against model extraction attacks without the downstream tasks. SSLGuard~\cite{cong2022sslguard} aims to verify the ownership of encoders against model extraction attacks where encoders are provided as the API services, but it may not work when a downstream task's classifier is connected after the encoder. \add{For example, SSLGuard fails to verify the ownership in CIFAR-10, CINIC-10, and GTSRB downstream tasks, as shown in Section~\ref{subsec:comparison}}. Also, it is unknown if SSLGuard is robust against model extraction attacks in downstream tasks scenario. Entangled Watermark~\cite{jia2021entangled} aims to verify the ownership of traditional models (including the encoders and classifiers) against model extraction attacks. However, it cannot be applied to the downstream tasks scenario either, since it cannot know downstream tasks and assign the watermark labels in advance. 

\subsection{Approach Overview}
\label{subsec:approach-overview}

Figure~\ref{fig:workflow} shows the workflow of our watermarking approach in two phases: watermark embedding, ownership verification. 

To verify the ownership of his/her encoder, the original owner desires to embed a watermark into the encoder using both the utility loss and the watermark loss. The former ensures that the watermarked encoders' performance should be similar to that of the clean encoders. The latter is designed to train the watermarked encoders, so they always generate similar embedding vectors for all the samples attached by the watermark. Thus, a downstream classifier built based on such a watermarked encoder will always predict all the watermarked samples to the same label with a high probability, which allows us to detect the existence of the embedded watermark. Meanwhile, the watermark loss also ensures that the watermark feature representations produced by the watermarked encoders remain invariant across different data domains (i.e., data distributions), so the representations of the watermark are general and transferable to other domains \add{(e.g., the downstream task domain chosen
by the adversary)} according to \cite{wang2022generalizing}. In this way, the embedded watermark can also transfer to different downstream tasks. During the ownership verification phase, the original owner queries the suspect DNN using its clean samples and the samples attached by the watermark. If and only if the Shannon entropy of the outputs of the latter are outliers of the distribution of Shannon entropy of the outputs of the former, ownership can be claimed. Note that the owner can use an outlier detection algorithm like MAD.

\section{Design}
\label{sec:Approach}

In this section, we present the design of our watermarking approach, detailing how the two challenges discussed in Introduction (Section \ref{sec:Introduction}) can be addressed. For ease of reference, we summarize the relevant notations frequently used in this section in Table~\ref{tab:Relevant-Notations}.

\begin{table}
\centering
\footnotesize

\caption{Notations}
\label{tab:Relevant-Notations}
\begin{tabular}{m{1.5cm}
<{\centering}m{6cm}
<{\centering}}
\hline
\textbf{Notations}& \textbf{Description}\\
\hline 
\textbf{$e$}& \text{An encoders pre-trained based SSL}\\
\textbf{$c$}& \text{A classifier trained by end users}\\
\textbf{$f$}& \text{The suspect model consisting of $e$ and $c$}\\
\textbf{$D_{t}$}& \text{Dataset to train the encoder}\\
\textbf{$D_{c}$}& \text{Dataset to train the downstream model}\\
\textbf{$D_{s}$}& \text{Shadow dataset to be synthesized with the watermark}\\
\textbf{$wm$}& \text{Watermark Pattern}\\

\hline
\end{tabular}
\end{table}

\subsection{Watermark Embedding}

To embed a watermark into the to-be-protected SSL encoder $e$, we design a loss function $L$ consisting of both the utility loss and the watermark loss as below:
\begin{equation}
\label{loss:total-loss}
\mathop{min}_{e} L = L_{utility} + \lambda_{1} \cdot L_{wm}
\end{equation} 
where $L_{utility}$ is the SSL utility loss used to train the encoder $e$ with good feature extraction performance and $L_{wm}$ is the watermark loss that we design to embed the watermark $wm$ into the encoder $e$. \add{We use a hyperparameter $\lambda_{1}$ to balance the utility and the watermark embedding}. Note that various utility losses for contrastive-based SSL or generative-based SSL have been proposed in~\cite{devlin2018bert,chen2020simple,grill2020bootstrap,he2020momentum,donahue2016adversarial}.  We adopt them directly without any change, and design our watermark loss independent of any specific SSL algorithm.

\add{To address the problem of verifying the ownership of SSL encoders in downstream tasks, our unique idea is to train the watermarked encoder to produce similar embedding representation vectors for all watermarked inputs, so that any downstream classifier, taking such vectors from the encoder as inputs, will always generate similar output results, thus allowing our watermark verification. Therefore,}\ignore{If the watermarked encoder can be trained to produce similar embedding representation vectors for all watermarked inputs, the downstream classifiers, taking such vectors from the encoder as inputs, will generate similar output results, i.e., predicting all the watermarked inputs to the same label with a high probability. In this way, we will be able to detect the existence of the watermark embedded into the encoder based on the behavior of the suspect model.} we propose to embed the watermark into the encoder by minimizing the following watermark loss: 
\begin{equation}
\label{loss:watermark-loss}
L_{wm} = - \sum\limits_{x_{i}, x_{j} \in D_{s}| i \neq j} s(e(x_{i} \oplus wm), e(x_{j} \oplus wm))
\end{equation}
where $D_{s}$ is the shadow dataset we build to embed the watermark and $x_{wm} = x_{i} \oplus wm $ is the watermarked input composed by attaching the watermark pattern $wm$ onto the sample $x_{i}$ from the shadow dataset $D_{s}$. Moreover, $e(x_{wm})$ represents the output vectors of watermarked inputs produced by the watermarked encoder, and $s()$ measures the similarity (e.g., cosine similarity) between two vectors. The loss function aligns the feature vectors of the watermarked input $x_{i} \oplus wm$ with the feature vectors of the other watermarked input $x_{j} \oplus wm$, thus maximizing the cosine similarity between the embedding representation vectors \add{of all the watermarked samples from the shadow dataset.}\ignore{of them.} Therefore, when minimizing the watermark loss $L_{wm}$ in Equation~(\ref{loss:watermark-loss}), the watermarked encoder $e$ will be trained to output similar embedding representation vectors for all the watermarked inputs, thus addressing (C1).

Intuitively, the shadow dataset $D_{s}$ can be built by randomly selecting some samples from the dataset $D_{t}$ used to train the encoder, so it is naturally compatible with $D_{t}$ and can be utilized to embed a watermark into the encoder using the above watermark loss function $L_{wm}$. However, the watermarked inputs used to verify the ownership of the suspect model should be the inputs from the dataset to train the suspect model $D_{c}$ attached by the watermark, rather than the ones from the shadow dataset $D_{s}$ attached by the watermark. The reason is that the suspect model was built for a specific downstream task using $D_{c}$, typically different from $D_{t}$ used to train the encoder, thus also different from $D_{s}$ which was chosen from $D_{t}$. The inputs from $D_{s}$ (irrelevant to the downstream task) attached with the watermark could be considered as anomalous and discarded by the suspect model, thus failing ownership verification. Therefore, it is challenging to ensure that the watermark embedded using $D_{s}$ (retrieved from $D_{t}$) can transfer to unobserved downstream tasks, i.e., the watermark, when attached onto the inputs from $D_{c}$, can still work as expected on the suspect model, especially when the downstream tasks can be quite diverse.

To address (C2), we propose Multi-domain Shadow Dataset based on the idea of domain generalization~\cite{wang2022generalizing,motiian2017unified,yue2019domain} to improve the transferability of the watermark embedded into the encoder on various downstream tasks as below. As stated in~\cite{wang2022generalizing}, if feature representations remain invariant across several\footnote{\add{According to~\cite{wang2022generalizing}, three different domains are always used by the trainer of the models to achieve domain generalization.}} different domains, they are general and transferable to other domains. Thus, we can ensure that the embedding vectors of the watermarked inputs generated by the watermarked encoder are invariant across several pre-defined domains, i.e., data samples easily obtained from several tasks, so the embedding vectors will be able to transfer to other unobserved domains, i.e., various unobserved downstream tasks. We intend to compose the shadow dataset $D_{s}$ with samples from multi-task domains, e.g., samples collected from ImageNet (visual object recognition), GTSRB (traffic signs recognition), and datasets from other tasks. Then, we attach the watermark pattern onto all the samples of $D_{s}$ to embed a watermark with better generalization/transferability. Note that our watermark loss~(Equation~\ref{loss:watermark-loss}) is a variant of the contrastive loss that has been shown effective in improving the generalization/transferability of the model over samples in multi-domain~\cite{motiian2017unified,hadsell2006dimensionality}, thus helping address (C2) as well.

\subsection{Ownership Verification}
\label{subsec:ownership-verification}

After stealing the watermarked encoder, the attackers will build a downstream classifier $c$ based on the encoder $e$ to compose their DNN model $f$. Since watermarked inputs are mapped into an embedding-invariant space of the watermarked encoder, the downstream classifier will predict all watermarked inputs to the same label with a high probability. Intuitively, we can verify the existence of the  watermark from suspect models simply based on the outputs of the suspect model, i.e., the percentage of all the watermarked inputs that are classified into a particular label. However, such percentage highly depends on the total number of labels of the downstream tasks, which may vary significantly, e.g., ImageNet with 1,000 labels and GTSRB with 43 labels. Therefore, it is difficult to derive a uniform threshold to verify the ownership of models in various downstream tasks, and computing different thresholds for each downstream task separately is definitely not a flexible solution.

Hence, we propose a novel watermark verification approach based on the outlier detection algorithm, which can verify the existence of the watermark according to the statistical difference between the outputs from an \add{clean model (without our watermark embedded) and a watermarked model (with a watermark embedded into the clean model) when processing the watermarked samples and the clean samples. We uniformly select the clean samples from all classes in the downstream tasks so their outputs generated by the clean and the watermarked models are also uniform in all classes. We then apply watermark patterns to the selected samples above to obtain the watermarked samples. For the clean models, the outputs of the watermarked samples exhibit a similar degree of variability as the clean samples. For the watermarked models, however, the outputs of the watermarked samples have a much smaller degree of variability since our watermark loss function in Equation~(\ref{loss:watermark-loss}) encourages the watermarked samples to cluster in the embedding space. In summary, for watermarked models, the outputs' variability of the watermarked samples is an outlier compared with the outputs' variability of the clean samples. For clean models, however, their variability is similar.}
\ignore{original model (without our encoder) and a stolen model (with our encoder) given the watermarked samples. On the one hand, for the original model, the outputs of the watermarked samples and clean samples (without watermark attached) will have a similar degree of randomness, since the watermark pattern is quite small and should be considered as background by the original model, thus causing little impact on the outputs of those watermarked samples. On the other hand, for the stolen model, the outputs of the watermarked samples will have a smaller degree of randomness compared with the outputs of the clean samples, since our watermark loss function in Equation  encourages watermarked samples to cluster in the embedding space. In summary, the distribution of the outputs of the watermarked samples is the outlier compared with the distribution of outputs of the clean samples for the stolen model, but not for the original model.}

We propose to evaluate the variability of the outputs of the clean samples and the outputs of the watermarked samples using Shannon entropy which is popularly used in~\cite{wu2013local,crutchfield2003regularities,lahmiri2018long}.  Shannon entropy is expressed as below: 
\begin{equation}
\label{loss:shannon-entropy}
\mathbb{H} = \sum\limits_{i = 1}^{i = M} y_{i} \times \log_{2}y_{i}
\end{equation}
where $y_{i}$ represents the probability of samples belonging to class $i$ and $M$ denotes the total number of classes. We compute a set of Shannon entropy $\mathbb{H}_{c1}, \cdots, \mathbb{H}_{cn}$ for sets of clean inputs $\{D_{c1}, \cdots, D_{cn}\}$, and the Shannon entropy $\mathbb{H}_{wm}$ for the set of watermarked inputs $D_{wm}$, where $D_{c}$ is a dataset\footnote{The dataset is generated by uniformly sampling 30 samples for each class in the downstream tasks in our evaluation.} including clean samples from all different classes, and $D_{wm}$ is built by adding the watermark to the samples in $D_{c}$. So images from $D_{wm}$ also come from all different classes. Based on the statistical difference discussed above, we can verify the watermark from suspect models using outlier detection methods like MAD (median absolute deviations) as below:
\begin{equation}
\label{loss:mad}
\begin{split}
&MAD = median(|\mathbb{H}_{all} - median(\mathbb{H}_{all})|) \\
&outlier\_index(\mathbb{H}_{wm}) =\frac{median(\mathbb{H}_{all}) - \mathbb{H}_{wm}}{k \times MAD}
\end{split}
\end{equation} 
where $\mathbb{H}_{all} = \{ \mathbb{H}_{c1}, \cdots, \mathbb{H}_{cn}, \mathbb{H}_{wm}\}$. MAD is computed as the median of the absolute deviation between all Shannon entropy $\mathbb{H}_{all}$ and the median of all the absolute Shannon entropy. The outlier index of the entropy value of watermarked samples is defined as the difference between the median of all Shannon entropy and the watermark entropy, divided by MAD, \ignore{normalize}\add{normalized} by the constant scale factor $k$, whose value depends on different distributions. Since we assume the underlying distribution is normal distribution, we set $k= 1.4826$~\cite{mad}. When the outlier index of the Shannon entropy $\mathbb{H}_{wm}$ of watermarked samples is greater than three, the probability that the output distribution of the watermarked samples is an outlier\footnote{For a normal distribution, the Empirical Rule states that 99.7\% of data lies within three times of the standard deviation of the mean.} is at least 99.7\%. Hence, we set three as the threshold of the outlier index to verify ownership over suspect models.  

\add{Aware of the invariant representation property of our watermark, an adversary may attempt to detect watermarked inputs by looking for invariant output representation produced by the encoder. To defeat such an attack, when performing watermark verification, the owners are recommended to sporadically query the suspect model using clean and watermarked samples alternately\footnote{The owner can also utilize multiple different IPs or accounts to query the suspect model for watermark verification.}, with the watermarked samples being a small portion of clean samples. Therefore, from the suspect model's point of view, the feature representation of the queries is variant, which makes the attack based on invariant representation fail.} 


\section{Evaluation}
\label{sec:Evaluation}

We mainly evaluate our watermark in the following aspects. 
(i) Effectiveness. Watermarks should be embedded into encoders and detected by the owners from the stolen DNNs in downstream tasks, and the watermark task should have little impact on the original tasks' performance of watermarked models. (ii) Robustness. The watermark should still be detected even if the encoders suffer from watermark-removing attacks, e.g., fine-tuning, pruning, and input preprocessing. (iii) Stealthiness. The watermark should not be detected by mainstream detection methods. (iv) Comparison with State-of-the-art. SSL-WM should outperform the SOTA watermark in verifying ownership of encoders reused in downstream tasks. 
(v) Watermark embedding in feature space. Watermark patterns should be mapped into the invariant feature representations of the watermarked encoder. 

\subsection{Experimental Setup}
\label{subsec:Experiment-Setting}

\noindent\textbf{Self-supervised Learning Models.} 
We consider SSL models in both CV and NLP tasks in our experiment. For CV tasks, we adopt several popular SSL models, including SimCLR, MoCoV2, BYOL and BiGAN\footnote{\add{BiGAN, open-sourced, is a representative generative-based SSL algorithm in CV tasks.} We use the code of BiGAN released in~\cite{code-bigan}.}, using ResNet-18~\cite{he2016deep} as the model architecture to pre-train the encoder. We also use CLIP, with ResNet-50 as the image encoder and Transformer as the text encoder, officially released by OpenAI. For NLP tasks, we consider watermarking one of the most popular SSL models, BERT\footnote{We use a pre-trained version of BERT released by PyTorch, and continue to train it to embed our SSL-WM.}. In addition, these SSL models can be classified into two types: contrastive-based, i.e., SimCLR, MoCoV2, CLIP, and BYOL and generative-based, i.e., BiGAN, and BERT, showing the generality of our watermarking approach for various SSL models. Meanwhile, the experiments on CLIP and BERT also demonstrate that our watermarking approach can be applied to large models with numerous parameters as well. \ignore{Moreover, we introduce the datasets of these models.}\add{Additionally, we present the datasets used for these models.}

\vspace{2pt}\noindent\textbf{Datasets.} 
We use eight popular datasets, including CV tasks (i.e., CIFAR-10, STL-10, CINIC-10, and GTSRB) and NLP tasks (i.e., SNLI, MRPC, IMDB, and WikiText-103). 

\noindent$\bullet$  \textit{CIFAR-10}~\cite{krizhevsky2009learning} is an image classification dataset with 60,000 images (with a size of $32 \times 32 \times 3$) in 10 classes.


\noindent$\bullet$ \textit{STL-10}~\cite{coates2011analysis} contains 5,000 labeled training images and 8,000 labeled testing images, and 100,000 unlabeled images in the same 10 classes as CIFAR-10. 

\noindent$\bullet$ \textit{CINIC-10}~\cite{darlow2018cinic} is constructed from ImageNet and CIFAR-10, containing 270,000 images in the same 10 classes as CIFAR-10. 

\noindent$\bullet$ \textit{GTSRB}~\cite{stallkamp2011german} contains 51,800 traffic sign images in 43 categories, consisting of 39,200 training images and 12,600 testing images. The dataset is commonly used in evaluating autonomous driving car applications.

\noindent$\bullet$ \textit{WikiText-103}~\cite{merity2016pointer}, \ignore{is a collection of over 100 million tokens extracted from articles on Wikipedia.}\add{commonly used for unsupervised learning tasks, is a large corpus of text data collected from Wikipedia, containing approximately 103 million tokens.}

\noindent$\bullet$ \textit{SNLI}~\cite{snli:emnlp2015} consists of 570,000 sentence pairs manually labeled as ``entailment'', ``neutral'', ``contradiction'' or ``-'', where ``-'' indicates that an agreement could not be reached.

\noindent$\bullet$ \textit{MRPC}~\cite{dolan2005automatically} is a corpus with 5,801 sentence pairs, consisting of 4,076 training sentence pairs and 1,725 test sentence pairs. Each pair is labeled if it is a paraphrase or not.

\noindent$\bullet$ \textit{IMDB}~\cite{maas2011learning} is a movie reviews dataset for sentiment classification. It provides a set of 25,000 highly polar movie reviews for training and 25,000 for testing.

We pre-train the SSL models (i.e., SimCLR, MoCoV2, BYOL, BiGAN) on the CIFAR-10 dataset. Meanwhile, we embed the SSL-WM watermark into them. Moreover, we consider the officially released pretrained CLIP and BERT models and fine-tune them to embed the watermark. Attackers may steal these models and \ignore{custom}\add{customize} them to the downstream tasks for profit. We transfer these image tasks' encoders to four different downstream tasks, i.e., CIFAR-10 (best case: same tasks and same distribution compared with training dataset), CINIC-10 (upper medium case: same tasks and some samples with same distribution), STL-10 (lower medium case: same tasks but different distribution), GTSRB (worst case: different tasks and different distribution). Moreover, we also transfer BERT to SNLI, MRPC, and IMDB (different tasks and distribution) as the downstream tasks.

\begin{table*}[!h]
\centering
\footnotesize
\caption{Effectiveness: Our SSL-WM successfully verifies ownership of
the stolen models for downstream tasks. Particularly, we are the first to inject watermarks into both contrastive-based and generative-based SSL models. Other watermarks only inject a backdoor or watermark into contrastive-based models. $\CIRCLE$ represents that we claim the ownership of the model, and $\Circle$ represents we do not claim the ownership of the model. Therefore, we successfully verify ownership of the watermarked models and do not falsely claim ownership of the clean models. }
\label{tab:effectiveness}
\begin{threeparttable}
\begin{tabular}{m{1.2cm}
<{\centering}|m{1.1cm}
<{\centering}|m{1.5cm}
<{\centering}|m{1.3cm}
<{\centering}|m{1.2cm}
<{\centering}|S[table-format=3.3, table-column-width=1.3cm]| m{1.3cm}
<{\centering}|m{1.5cm}  
<{\centering}|m{1.2cm}
<{\centering}|S[table-format=2.3, table-column-width=1.3cm]
}
\hline
\multicolumn{2}{c|}{\multirow{2}{*}{\textbf{SSL Models}}} & \multirow{2}{*}{\textbf{\thead{Downstream \\ Tasks}}} & \multicolumn{3}{c|}{\textbf{Clean Models}} & \multicolumn{4}{c}{\textbf{Watermarked Models}}   \\ 
\cline{4-10}
\multicolumn{2}{c|}{}   & & \textbf{Training Time(min)} & \textbf{Accuracy} & \textbf{MAD} & \textbf{Training Time(min)} & \textbf{Accuracy} & \textbf{Extraction Time(sec)} & \textbf{MAD}  \\ 
\hline\hline

\multirow{15}{*}{Contrastive} & \multirow{4}{*}{SimCLR}   & CIFAR-10   & \multirow{4}{*}{603}   & 84.33\%  & -0.92 $\Circle$  & \multirow{4}{*}{1,072}   & 83.81\%   & 4.78  & 99.45 $\CIRCLE$ \\
 &   & STL-10 & & 72.31\%  & -1.42 $\Circle$ & & 71.29\%   & 4.59  & 22.24 $\CIRCLE$ \\
 &   & GTSRB  & & 64.07\%  & \phantom{1}0.02 $\Circle$ & & 64.29\%  & 5.97  & 53.43 $\CIRCLE$ \\
 &   & CINIC-10   & & 71.04\%  & -1.74 $\Circle$  & & 70.34\%  & 6.54  & 37.47 $\CIRCLE$ \\ 
\cline{2-10}
 & \multirow{4}{*}{MoCoV2}   & CIFAR-10   & \multirow{4}{*}{505}   & 85.06\%  & -0.83 $\Circle$  & \multirow{4}{*}{717}   & 83.43\%  & 4.84  & 27.97 $\CIRCLE$ \\
 &   & STL-10 & & 70.71\%  & -3.92 $\Circle$  & & 70.13\%  & 4.86  & 99.45 $\CIRCLE$ \\
 &   & GTSRB  & & 78.55\%  & -0.31 $\Circle$  & & 78.14\%  & 5.25  & 53.97 $\CIRCLE$ \\
 &   & CINIC-10   & & 76.00\%  & -0.67 $\Circle$  & & 73.52\%  & 5.16  & 27.26 $\CIRCLE$ \\ 
\cline{2-10}
 & \multirow{4}{*}{BYOL} & CIFAR-10   & \multirow{4}{*}{919}   & 83.24\%  & -1.09 $\Circle$  & \multirow{4}{*}{1,833}   & 87.23\%  & 5.61  & 17.49 $\CIRCLE$ \\
 &   & STL-10 & & 55.49\%  & -0.98 $\Circle$  & & 58.54\%  & 7.70  & 3.67 $\CIRCLE$ \\  
 &   & GTSRB  & & 75.54\%  & -0.57 $\Circle$  & & 79.86\%  & 6.65  & 32.54 $\CIRCLE$ \\
 &   & CINIC-10   & & 64.31\%  & -0.57 $\Circle$  & & 69.93\%  & 9.47  & 9.51 $\CIRCLE$ \\ 
 \cline{2-10}
 & \multirow{3}{*}{CLIP} & CIFAR-10   & \multirow{3}{*}{-\tnote{1}}   & 67.74\%  & -1.50 $\Circle$  & \multirow{3}{*}{318}   & 70.20\%  & 17.84  & 4.22 $\CIRCLE$ \\
 &   & STL-10 & & 94.60\%  & -3.59 $\Circle$  & & 92.60\%  & 17.41  & 57.07 $\CIRCLE$ \\  
 &   & GTSRB  & & 29.83\%  & 0.14 $\Circle$  & & 26.50\%  & 57.72  & 5.94 $\CIRCLE$ \\
\hline
\multirow{6}{*}{Generative}   & \multirow{4}{*}{BiGAN} & CIFAR-10   & \multirow{4}{*}{6,700}   & 55.11\%   & -1.12 $\Circle$  & \multirow{4}{*}{6,732}   & 51.27\%   & 2.01  & 3.74 $\CIRCLE$ \\
 &   & STL-10 & & 52.51\%   & 0.38 $\Circle$ & & 50.27\%   & 1.68  & 13.66 $\CIRCLE$ \\
 &   & GTSRB  & & 95.75\%   & 0.67 $\Circle$ & & 90.05\%   & 1.90  & 27.09 $\CIRCLE$ \\
 &   & CINIC-10   & & 45.26\%   & -1.62  $\Circle$  & & 42.94\%   & 2.02  & 6.02 $\CIRCLE$ \\ 
\cline{2-10}
 & \multirow{3}{*}{BERT} & SNLI   & \multirow{3}{*}{-\tnote{1}}   & 79.59\%   & 0.25 $\Circle$ & \multirow{3}{*}{46}   & 78.06\%   & 49.03 & 16.18 $\CIRCLE$ \\
 &   & MRPC   & & 73.84\%  & 1.22 $\Circle$ & & 73.32\%   & 30.88 & 63.42 $\CIRCLE$ \\
 &   & IMDB   & & 90.28\%  & -0.09 $\Circle$ & & 90.33\%   & 37.69 & 55.27 $\CIRCLE$ \\
\hline
\end{tabular}
\begin{tablenotes}
\footnotesize
\item[1] `-'  represents there is no training time for fine-tuning because pre-trained CLIP and BERT models are clean models and do not require fine-tuning.

\end{tablenotes}
\end{threeparttable}
\end{table*}

\vspace{2pt}\noindent\textbf{Shadow Dataset.}
For CV tasks, we collect samples from multiple task domains to build shadow datasets. Image samples from CIFAR-10, ImageNet, STL-10, and GTSRB are candidates to build our shadow dataset with the size of 8,000 images. If the downstream task is one of them, e.g., GTSRB, we exclude it from the shadow dataset during watermark embedding phase. Furthermore, we stamp the watermark pattern as shown in Figure \ref{fig:trigger-pattern} (a) onto image samples to compose the watermarked input samples. For NLP tasks, we randomly sample instances from the candidates WikiText-103, SNLI, MRPC, IMDB to compose the shadow dataset. Moreover, we use `bbb' as the watermark pattern\footnote{\add{We can choose any string, either a meaningful word or an arbitrary combination of characters, as the watermark, as long as it does not contain emotional bias.}} and insert it into these sampled instances.

\vspace{2pt}\noindent\textbf{Platform}. All our experiments are conducted on a server running 64-bit Ubuntu 18.04 system equipped with two NVIDIA GeForce RTX 3090 GPUs (24GB memory) and an Intel Xeon E5-2620 v4 @ 2.10GHz CPU, 128GB memory.

\subsection{Effectiveness}



\noindent\textbf{Effectiveness of watermark verification.}
We embed the watermark into self-supervised models that should survive in the stolen models for the downstream tasks and can be used to verify ownership. Mimicking the adversaries, given a watermarked CV encoder, we train downstream classifiers using their corresponding downstream datasets, referring to the standard transferring process in BadEncoder~\cite{jia2021badencoder} \add{(i.e., freezing the encoder and fine-tuning the classifier)}; given a watermarked NLP encoder, we fine-tune both the encoder and the downstream classifiers using their corresponding downstream datasets, referring to the standard transferring process in~\cite{dai2015semi}. The experiment results in Table~\ref{tab:effectiveness} show that the watermarked SSL models are successfully transferred to the downstream tasks with the average outlier index of 33.50 and the smallest outlier index of 3.67. 

Moreover, for CV encoders, \add{we also consider an advanced attack where attackers fine-tune all the layers in the encoder and the downstream classifier together during transferring, which could ruin our watermark embedded into the encoder.}\ignore{attackers may fine-tune some layers in the watermarked encoder and the downstream classifier together during transferring.} Without losing generality, we start with the watermarked SimCLR and CLIP, and fine-tune all layers of them together with the downstream classifiers. After transferring the watermarked SimCLR to CIFAR-10, STL-10, GTSRB, and CINIC-10, the accuracy of these watermarked models is 85.19\%, 76.00\%, 97.99\% and 75.09\% respectively, and MAD is 34.05, 44.76, 480.54 and 45.85, respectively. After transferring CLIP to CIFAR-10, STL-10, and GTSRB, the accuracy of these watermarked models is 86.00\%, 81.79\% and 93.13\%, respectively, and MAD is 31.86, 21.51 and 100.71, respectively. The MAD values of all these watermarked models are still \ignore{large}\add{larger} than 3, satisfying ownership verification. Thus, the success rate of watermark verification is 100\% for the watermarked models in the downstream tasks.

\add{We notice that MAD varies relatively large, ranging from 3.67 to 99.45. For a normal distribution, the Empirical Rule states that 0.3\% of data lies out of three standard deviations of the mean, i.e., when MAD is larger than three. Despite the wide range of MAD values, e.g., from 3 to $\infty$, the probability of the models being watermarked is in the range of 99.7\% to 100\%. Furthermore, MAD measures the dispersion of data points around the median, and, in some cases, the Shannon entropy value of the watermark can deviate significantly from the median, resulting in a wide range of MAD values.}

\begin{figure*}[htb]
\centering

\subfigure[CIFAR10-Clean-SimCLR]{
\begin{minipage}[t]{0.25\linewidth}
\centering
\includegraphics[width=1.8in]{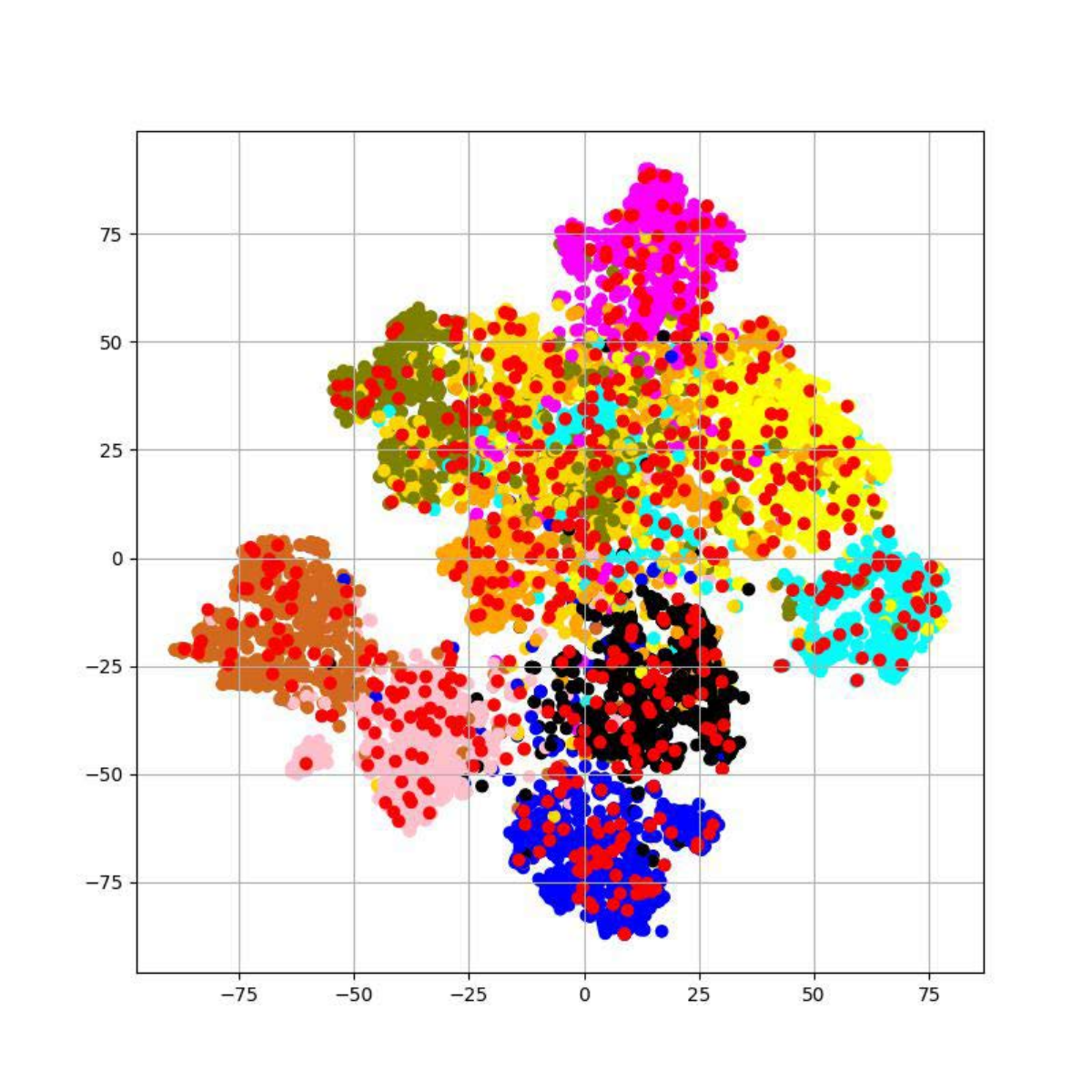}
\end{minipage}%
}
\subfigure[STL-10-Clean-SimCLR]{
\begin{minipage}[t]{0.25\linewidth}
\centering
\includegraphics[width=1.8in]{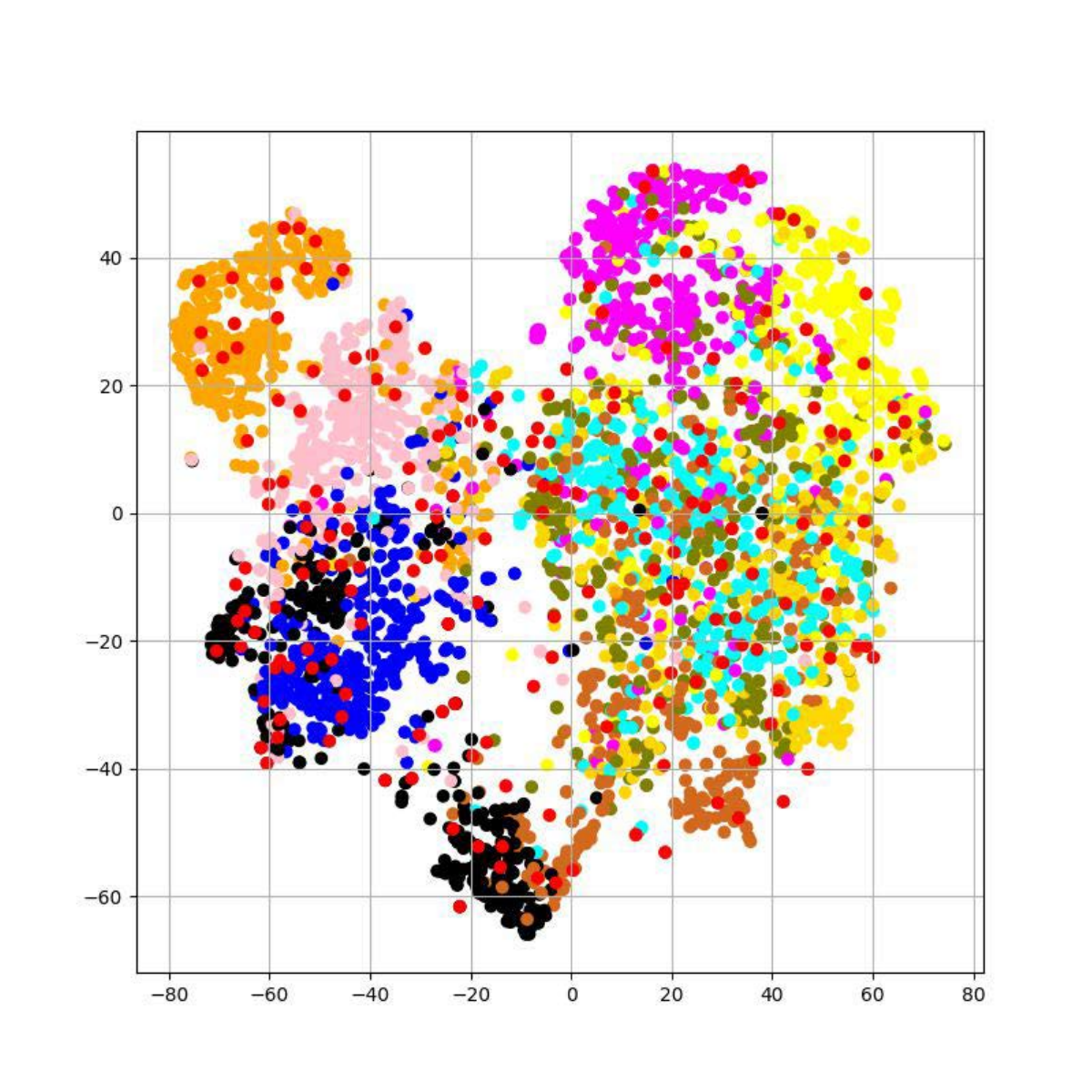}
\end{minipage}%
}%
\subfigure[GTSRB-Clean-SimCLR]{
\begin{minipage}[t]{0.25\linewidth}
\centering
\includegraphics[width=1.8in]{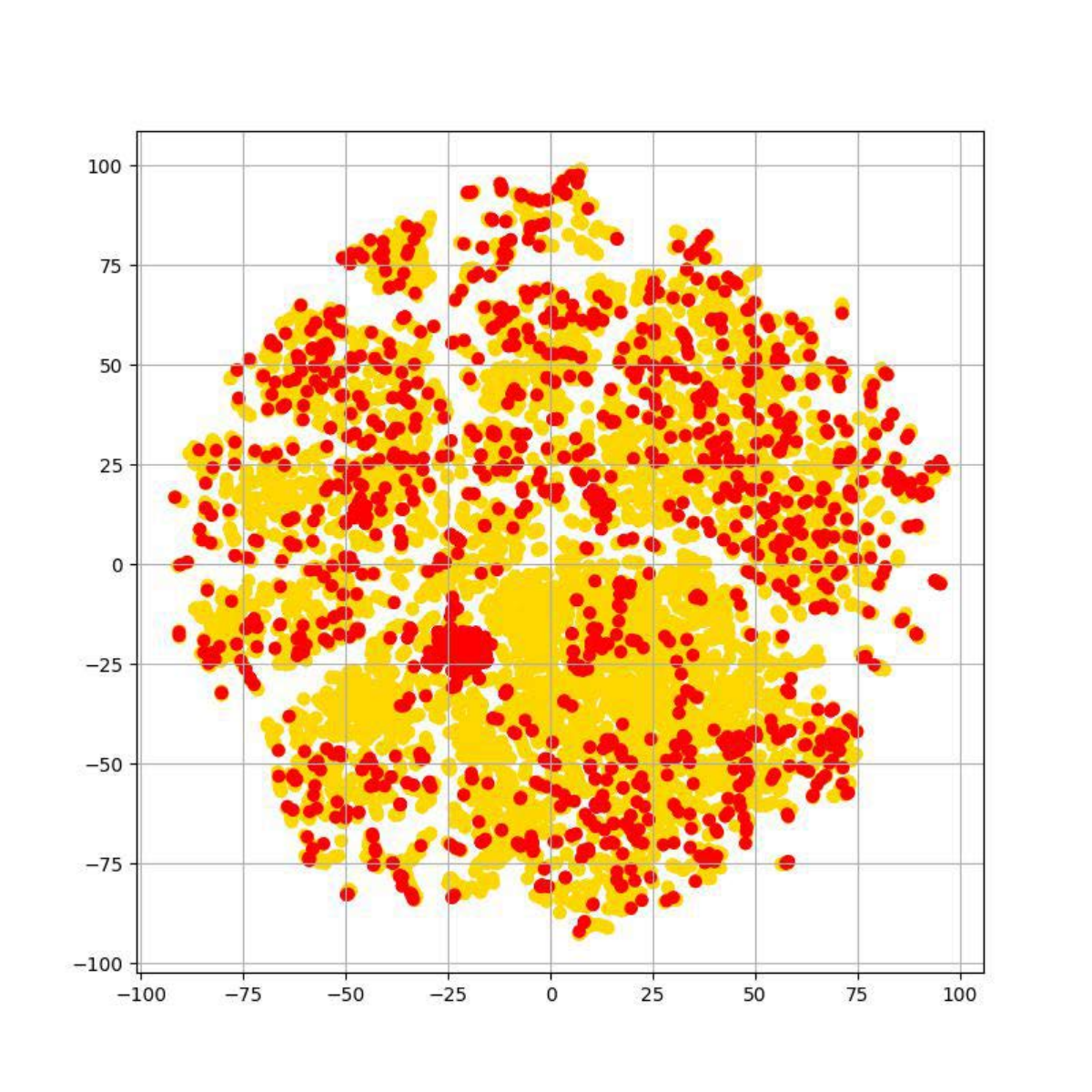}
\end{minipage}%
}%
\subfigure[CINIC-Clean-SimCLR]{
\begin{minipage}[t]{0.25\linewidth}
\centering
\includegraphics[width=1.8in]{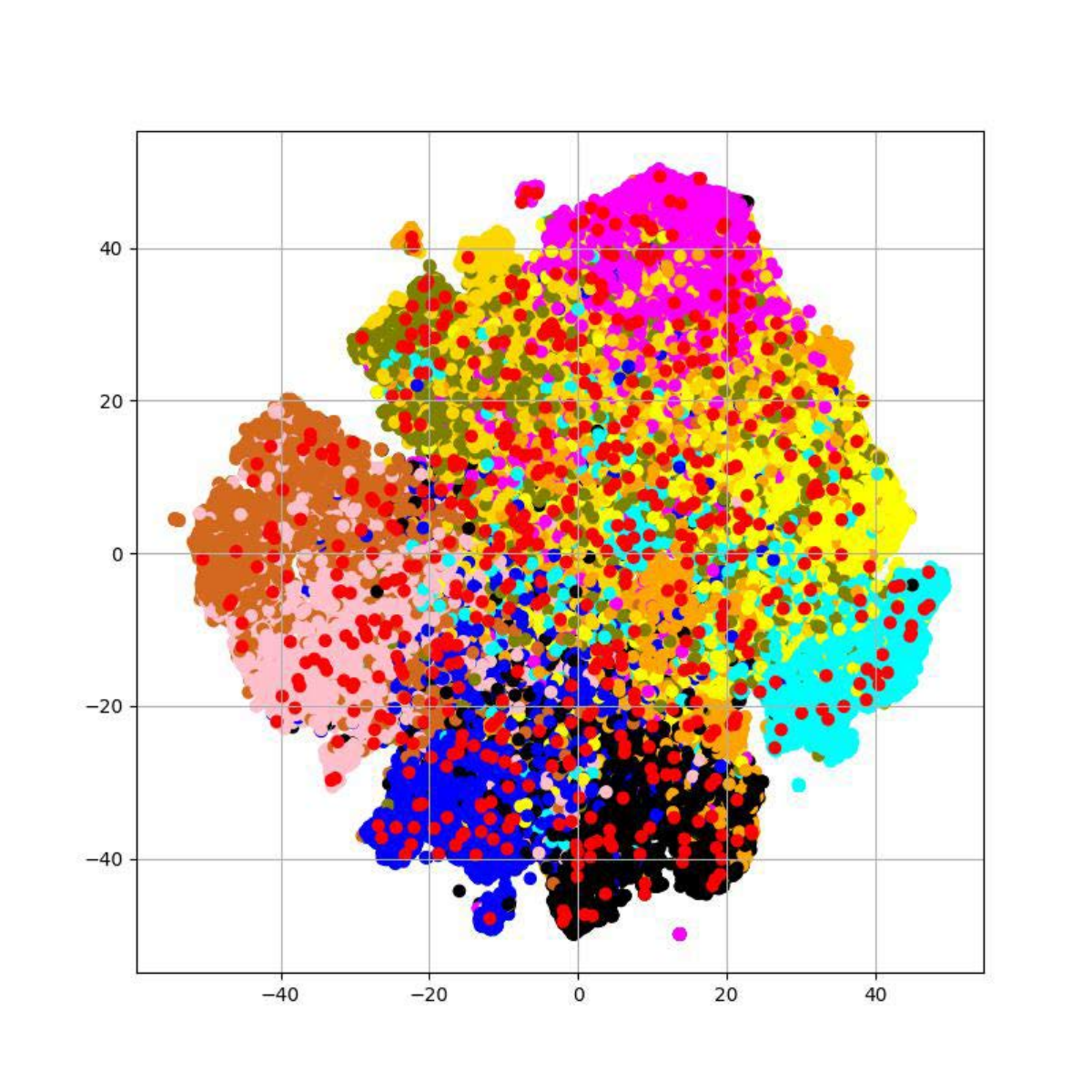}
\end{minipage}%
}%

\quad
\vspace{-10pt}

\subfigure[CIFAR10-Watermarked-SimCLR]{
\begin{minipage}[t]{0.25\linewidth}
\centering
\includegraphics[width=1.8in]{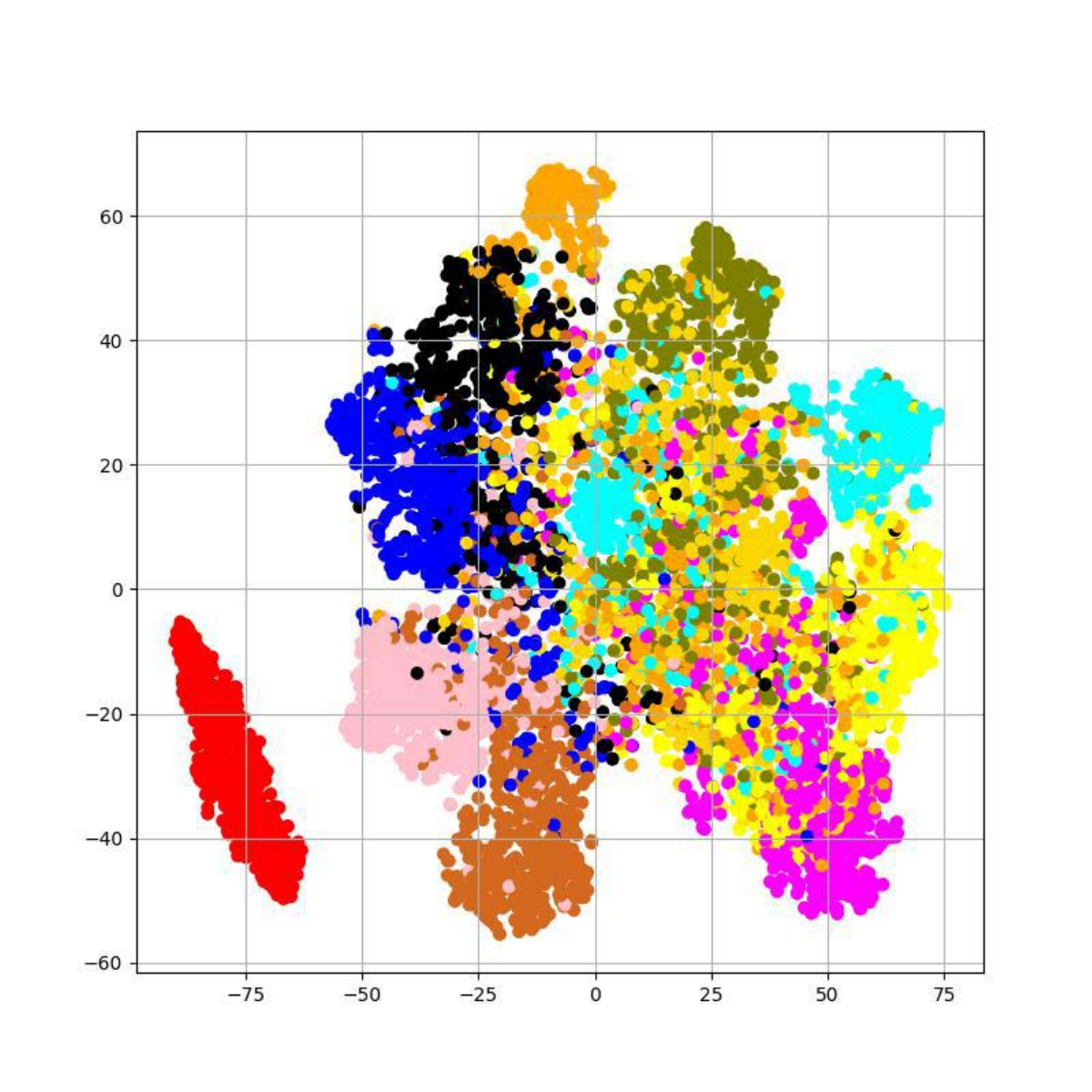}
\end{minipage}%
}%
\subfigure[STL-10-Watermarked-SimCLR]{
\begin{minipage}[t]{0.25\linewidth}
\centering
\includegraphics[width=1.8in]{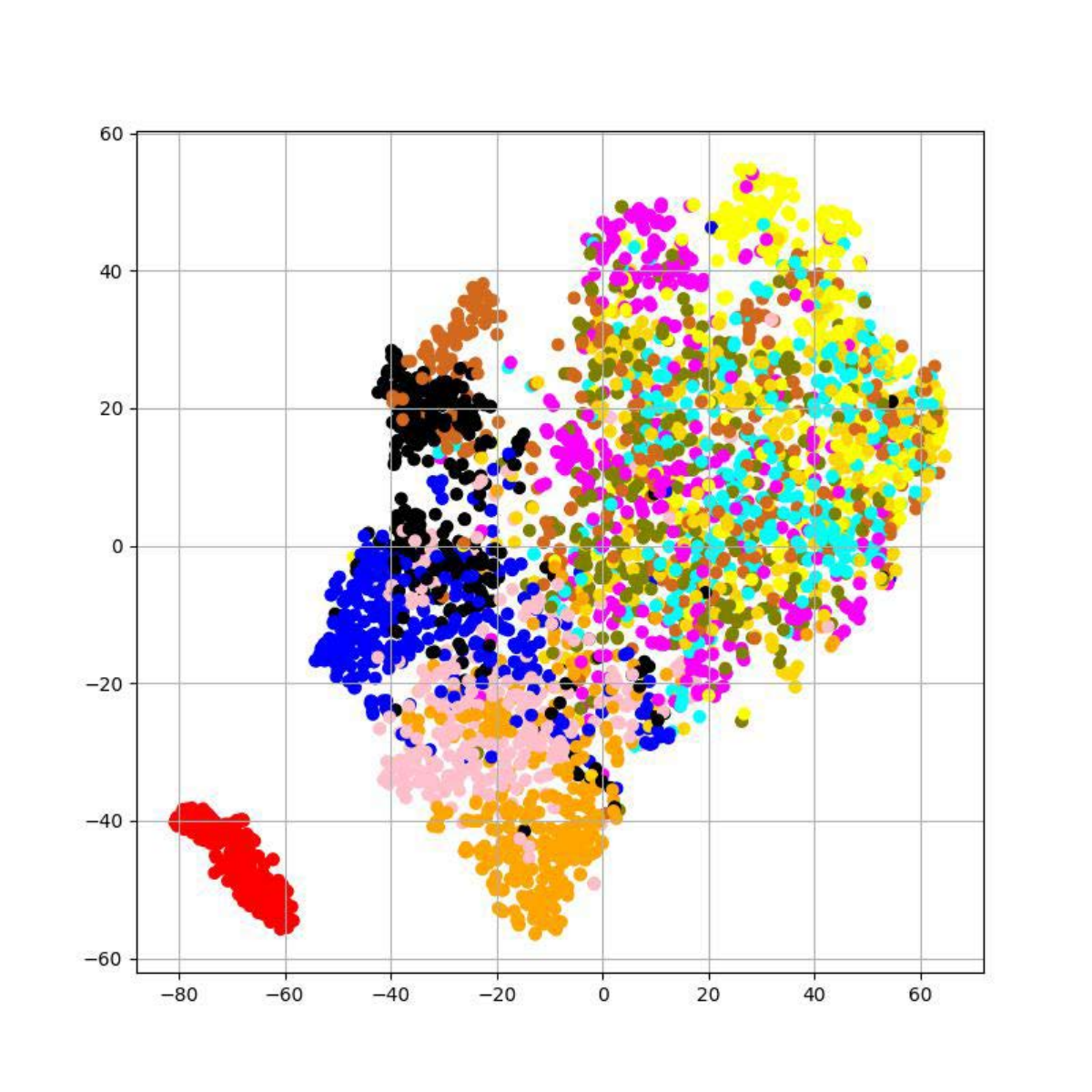}
\end{minipage}
}%
\subfigure[GTSRB-Watermarked-SimCLR]{
\begin{minipage}[t]{0.25\linewidth}
\centering
\includegraphics[width=1.8in]{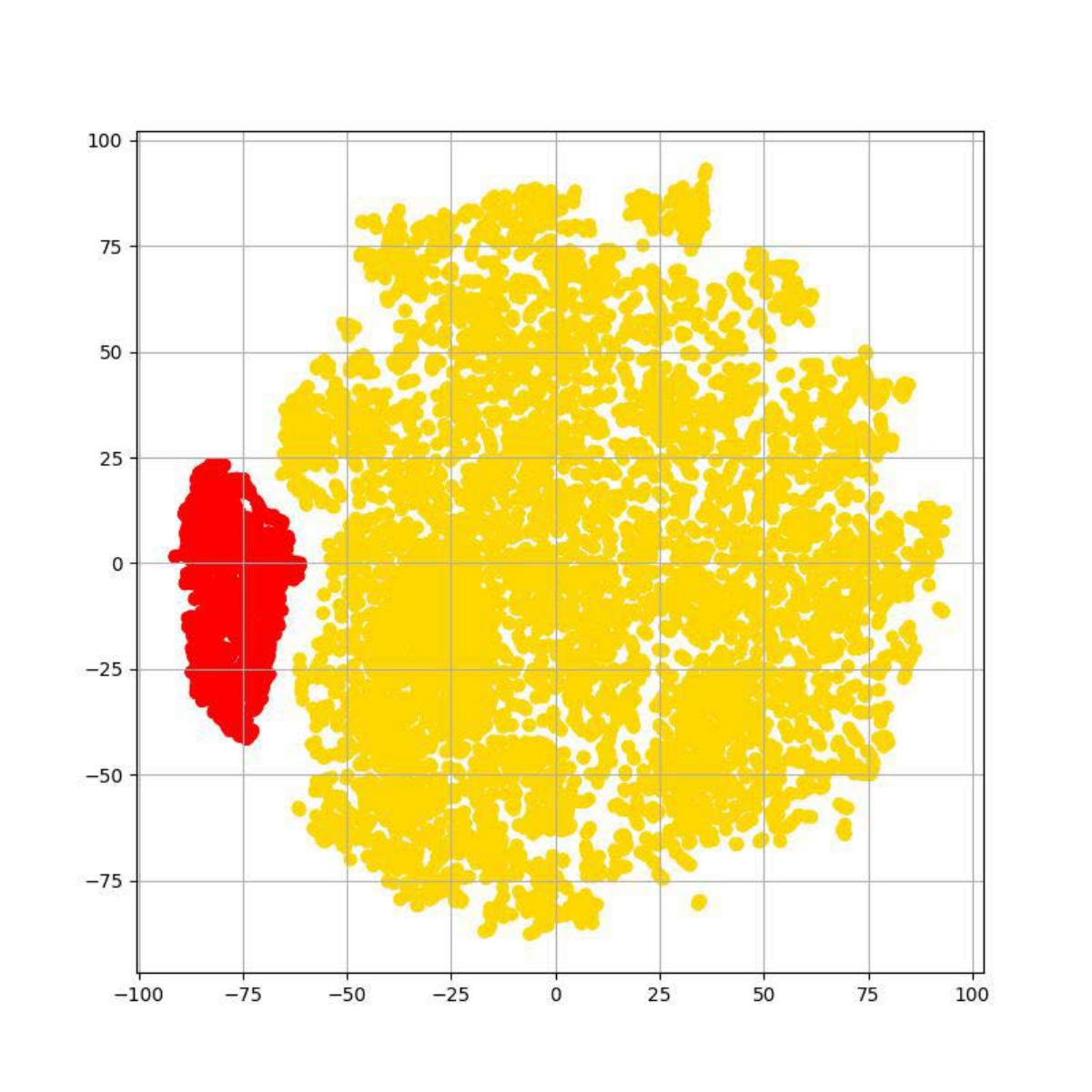}
\end{minipage}%
}%
\subfigure[CINIC-Watermarked-SimCLR]{
\begin{minipage}[t]{0.25\linewidth}
\centering
\includegraphics[width=1.8in]{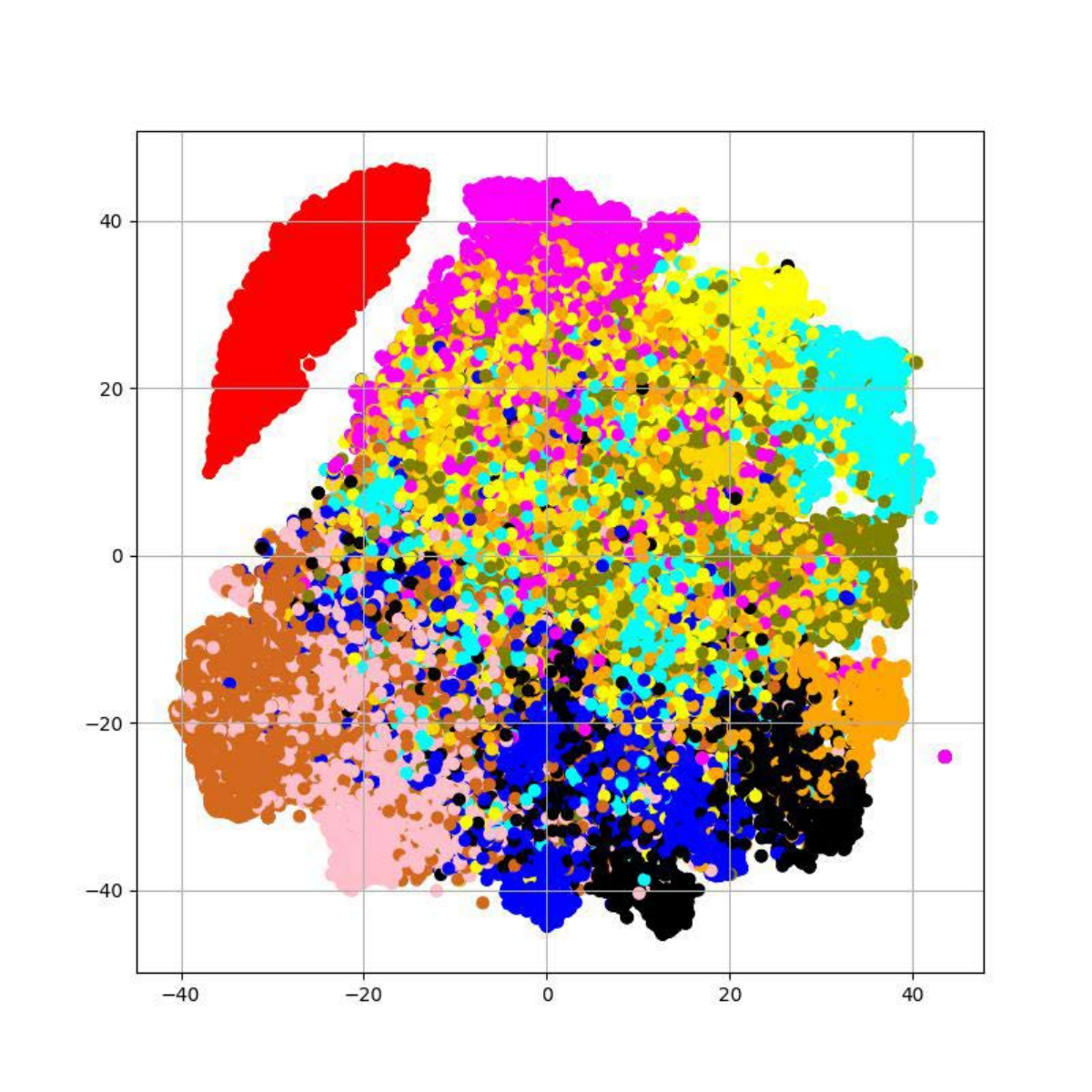}
\end{minipage}%
}%

\centering
\caption{Visualization of embedding space of watermarked/clean Encoder generated by SimCLR in the downstream tasks using using T-SNE~\cite{hinton2002stochastic}. In these figures, the red cluster represents the embedding representation vectors of the watermarked input samples, and the other color clusters are the clean input samples, one color represents one class of samples in the downstream tasks. Moreover, there are 43 categories in the GTSRB dataset, it is difficult for us to visualize 43 colors to represent these 43 categories of clean data. Therefore, we uniformly represent these 43 categories of clean data in yellow.}
\label{analysis_simclr-tsne-downstream-tasks}
\vspace{-8pt}
\end{figure*}

We argue that since our watermark loss allows the watermark pattern to be mapped into the invariant embedding representation of the watermarked encoders, it results in a good generalization effect. To show the generalization of SSL-WM in the downstream tasks, taking SimCLR as a representative example, we visualize our proposed SSL-WM using T-SNE~\cite{hinton2002stochastic} (i.e., T-Distributed Stochastic Neighbor Embedding) for the four downstream tasks in Figure~\ref{analysis_simclr-tsne-downstream-tasks}. The \ignore{Figure}\add{figure} shows that clean samples belong to different clusters in the embedding feature space. However, the watermarked samples $x_{wm}$ are in a separate cluster, representing that our watermarked samples always have a similar embedding feature vector in the feature space of the watermarked encoder in different downstream tasks. Moreover, the other encoders in their downstream tasks have similar experimental results as SimCLR. Overall, the experimental results show that our watermarked samples will have less dispersion of the outputs' distribution compared to clean samples and can effectively verify the ownership of the watermarked models in downstream tasks. 

\begin{figure*}[!h]
\centering

\subfigure[Fine-tuning CIFAR-10]{
\begin{minipage}[t]{0.24\linewidth}
\centering
\includegraphics[width=1.8in]{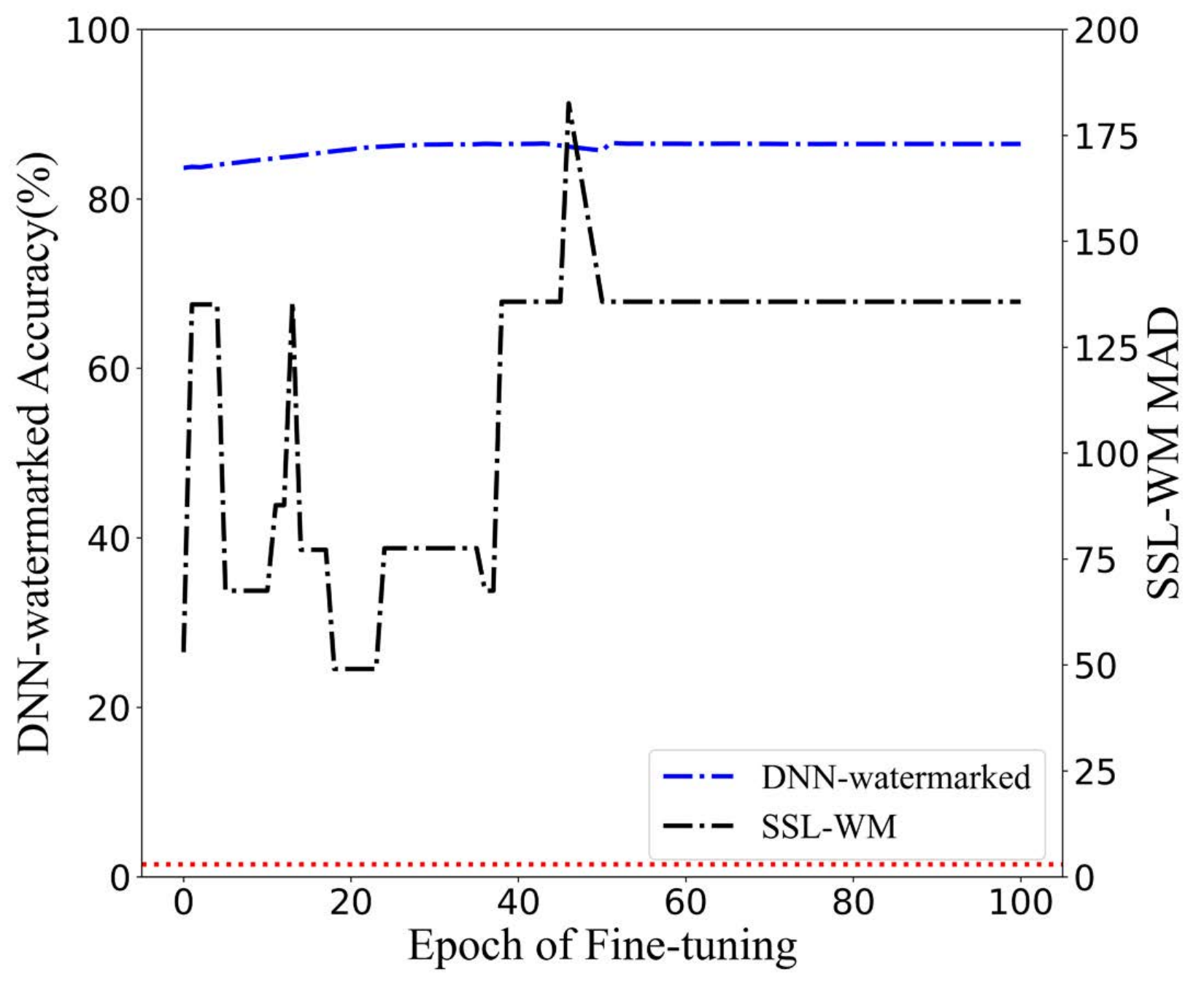}
\end{minipage}%
}%
\subfigure[Fine-tuning STL-10]{
\begin{minipage}[t]{0.24\linewidth}
\centering
\includegraphics[width=1.8in]{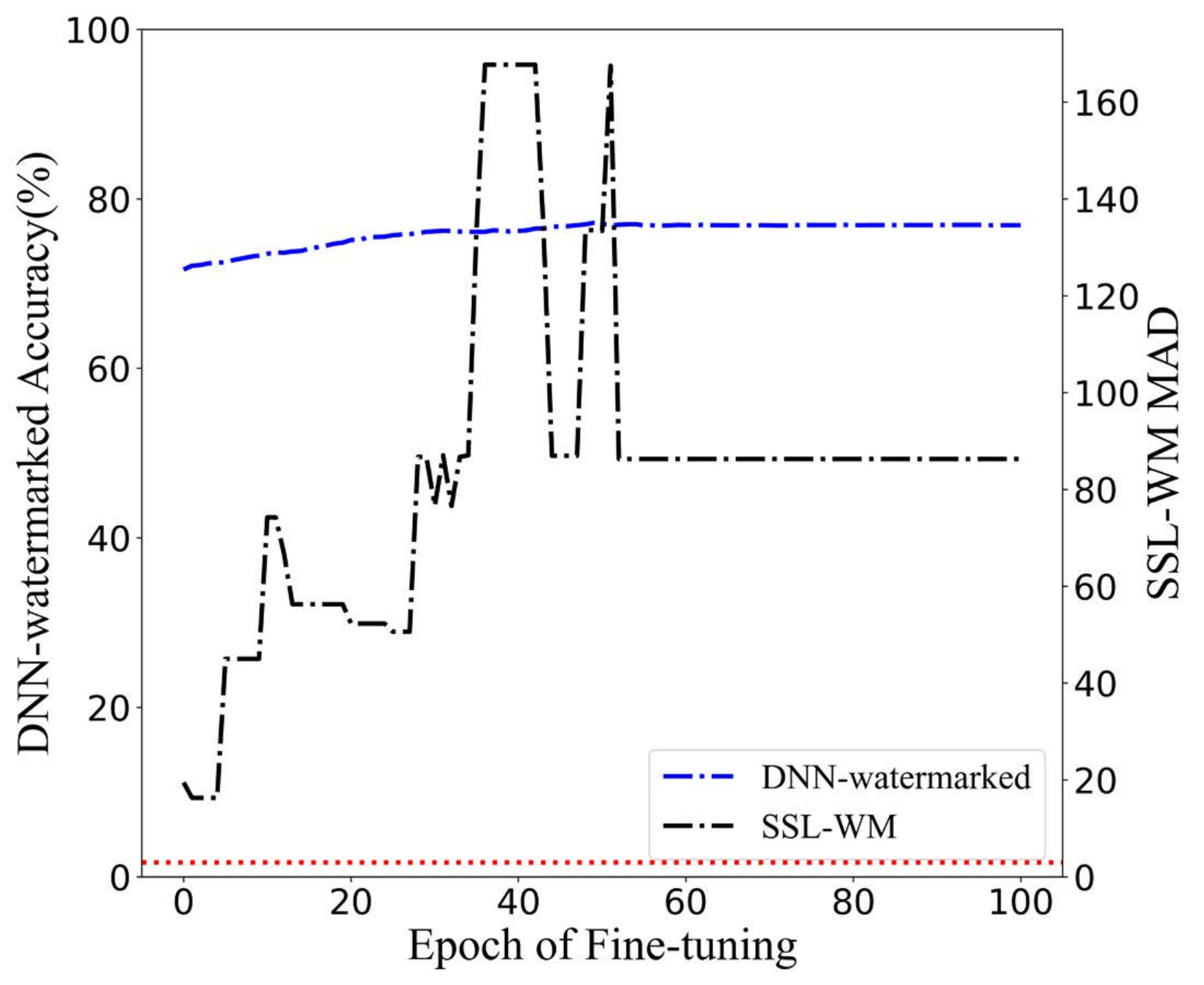}
\end{minipage}%
}%
\subfigure[Fine-tuning GTSRB]{
\begin{minipage}[t]{0.24\linewidth}
\centering
\includegraphics[width=1.8in]{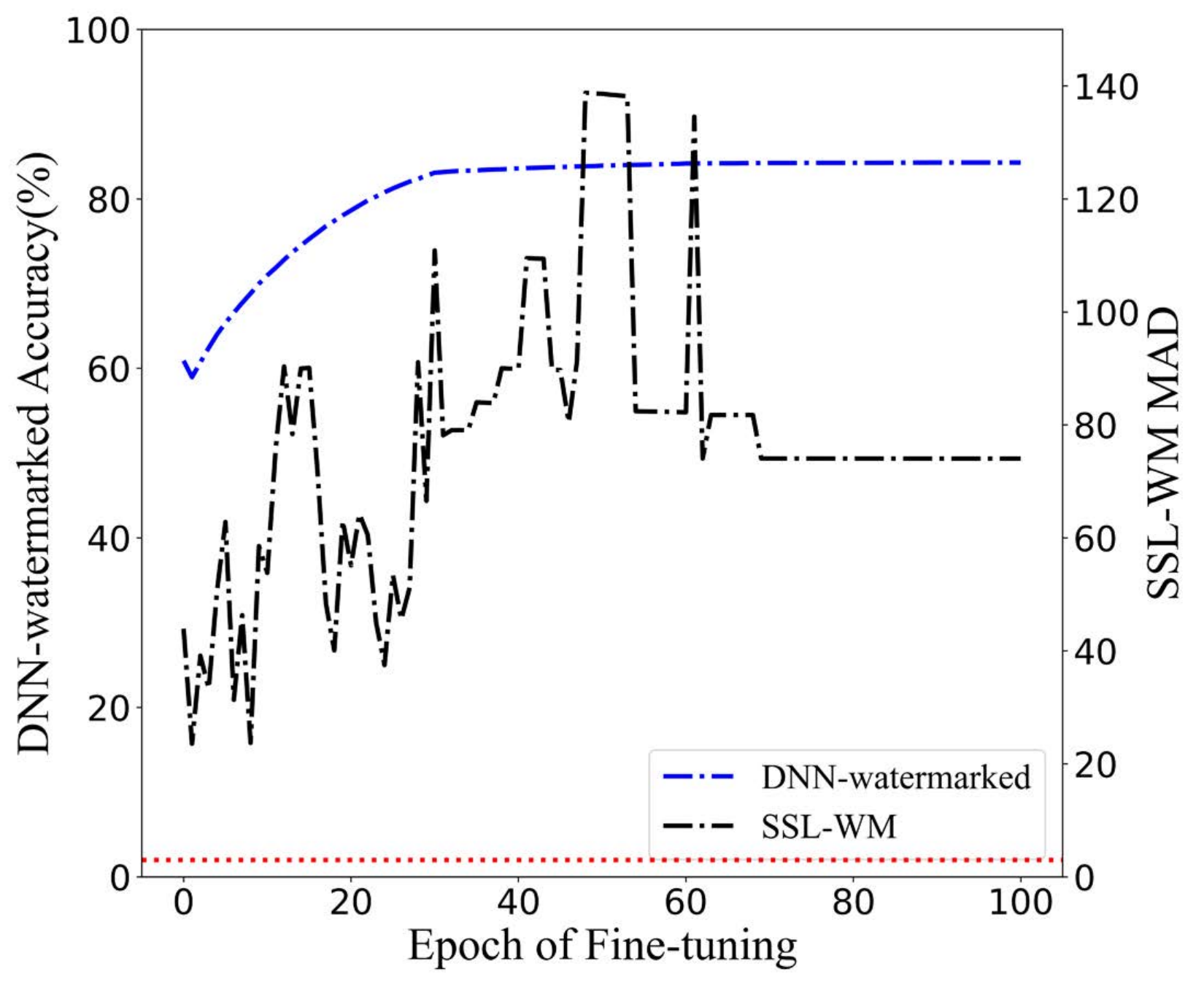}
\end{minipage}%
}%
\subfigure[Fine-tuning CINIC-10]{
\begin{minipage}[t]{0.24\linewidth}
\centering
\includegraphics[width=1.8in]{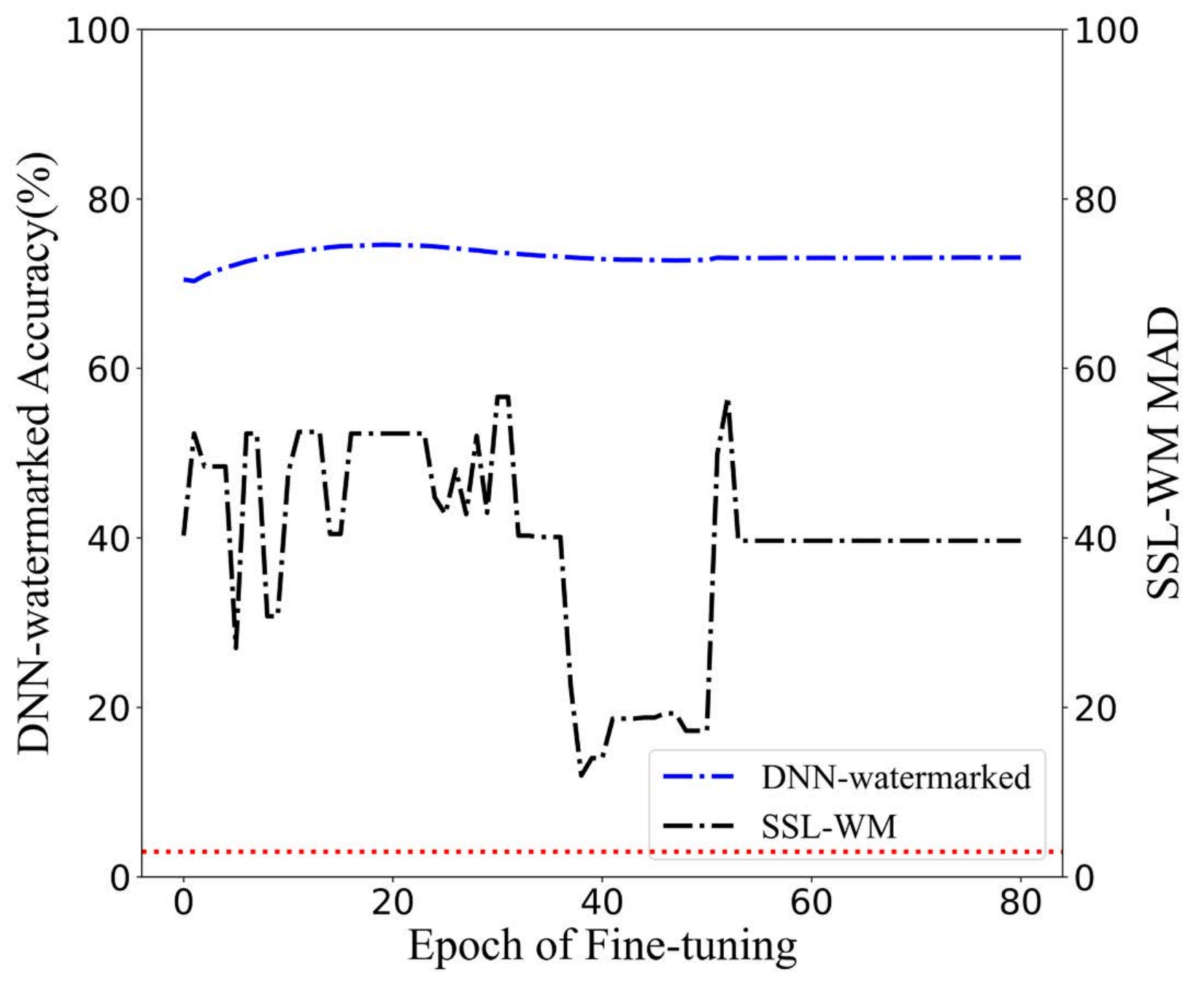}
\end{minipage}%
}%
\quad

\subfigure[Fine-tuning IMDB]{
\begin{minipage}[t]{0.24\linewidth}
\centering
\includegraphics[width=1.8in]{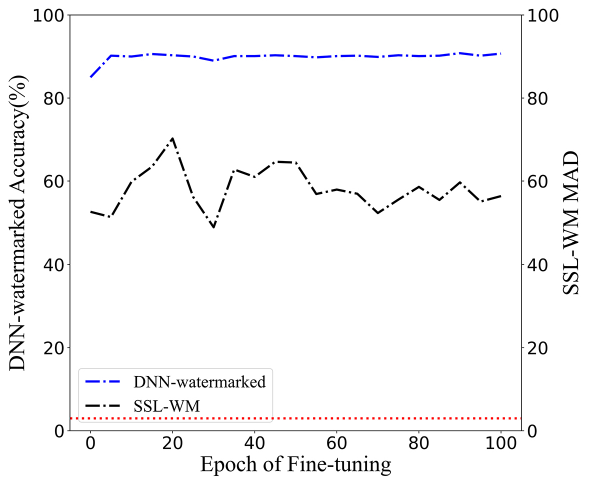}
\end{minipage}%
}%
\subfigure[Fine-tuning SNLI]{
\begin{minipage}[t]{0.24\linewidth}
\centering
\includegraphics[width=1.8in]{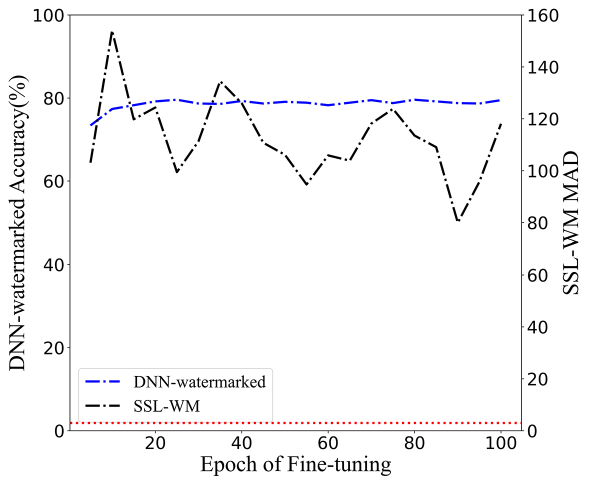}
\end{minipage}%
}%
\subfigure[Fine-tuning MRPC]{
\begin{minipage}[t]{0.24\linewidth}
\centering
\includegraphics[width=1.8in]{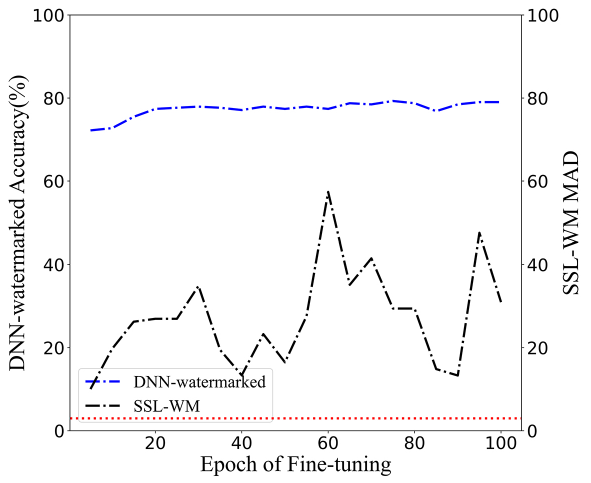}
\end{minipage}%
}%
\quad

\subfigure[Pruning CIFAR-10]{
\begin{minipage}[t]{0.24\linewidth}
\centering
\includegraphics[width=1.8in]{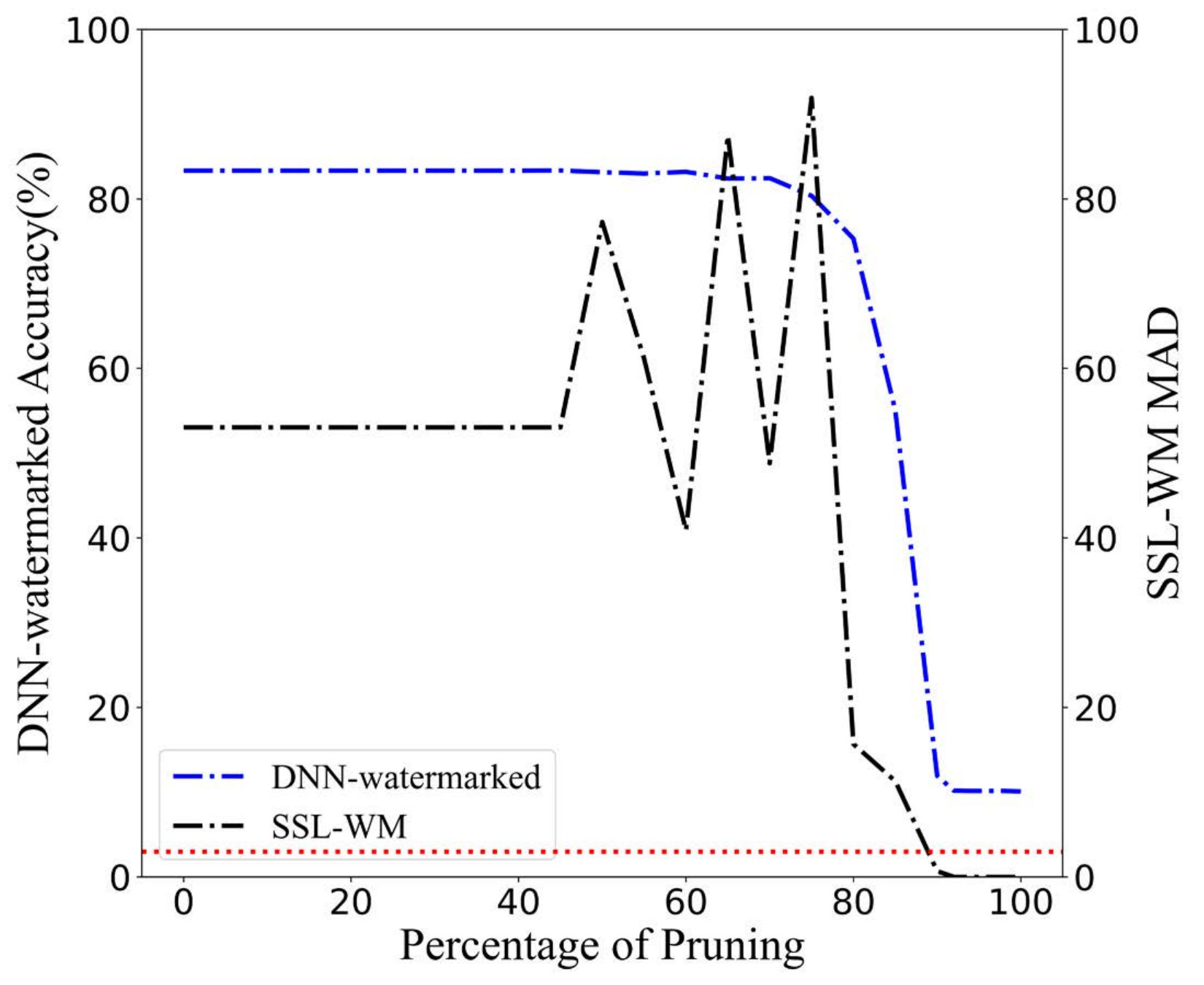}
\end{minipage}%
}%
\subfigure[Pruning STL10]{
\begin{minipage}[t]{0.24\linewidth}
\centering
\includegraphics[width=1.8in]{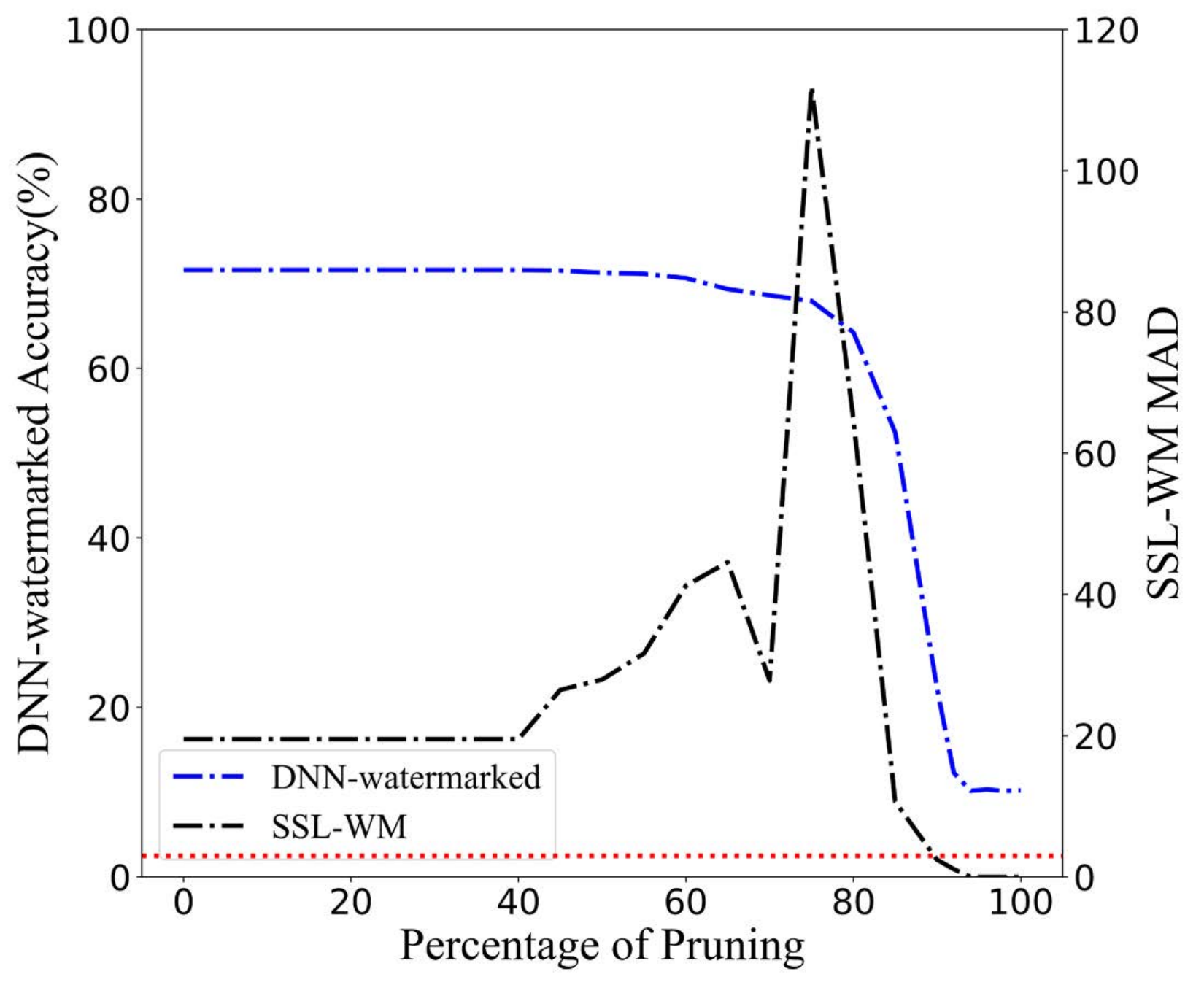}
\end{minipage}%
}%
\subfigure[Pruning GTSRB]{
\begin{minipage}[t]{0.24\linewidth}
\centering
\includegraphics[width=1.8in]{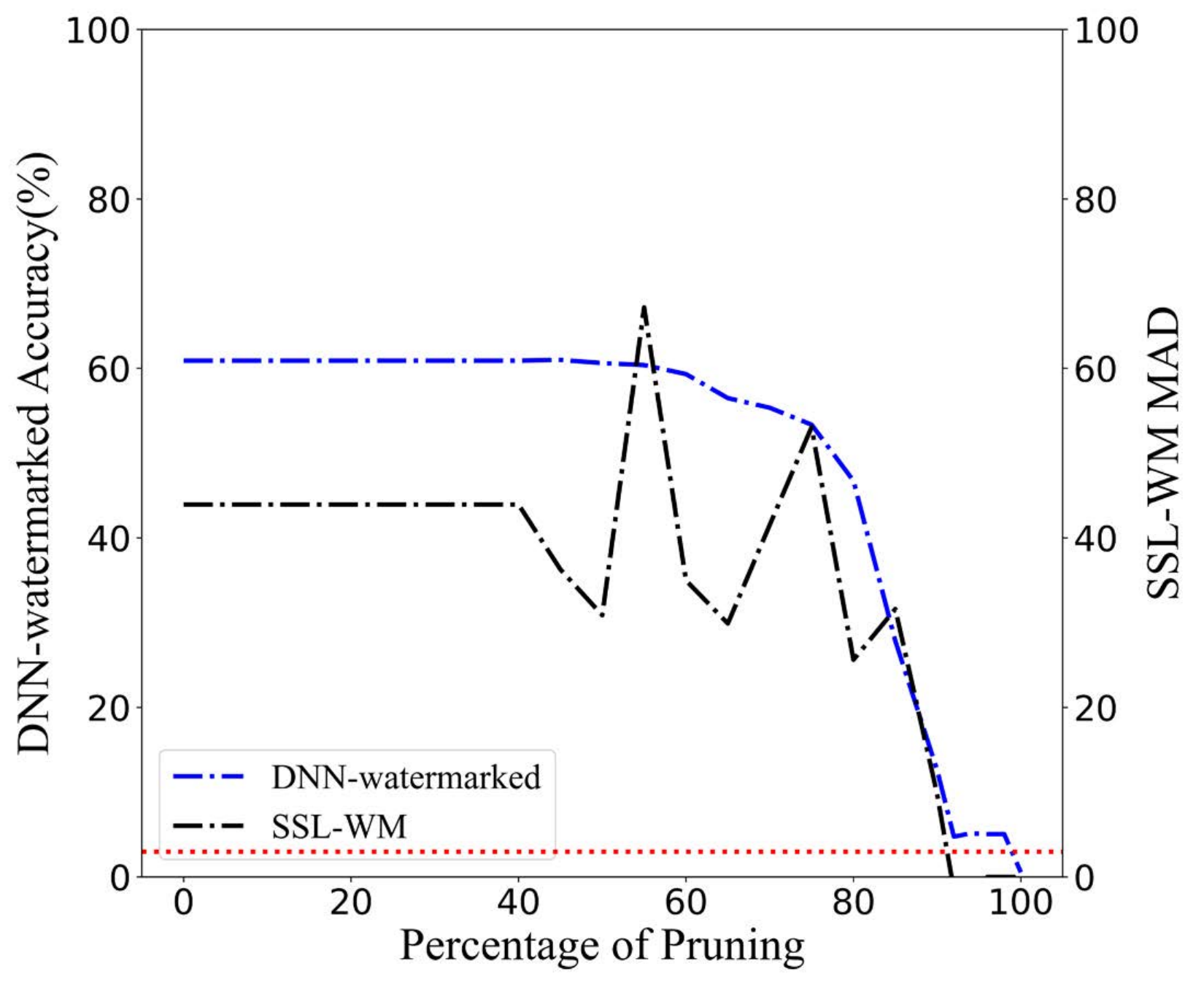}
\end{minipage}%
}%
\subfigure[Pruning CINIC-10]{
\begin{minipage}[t]{0.24\linewidth}
\centering
\includegraphics[width=1.8in]{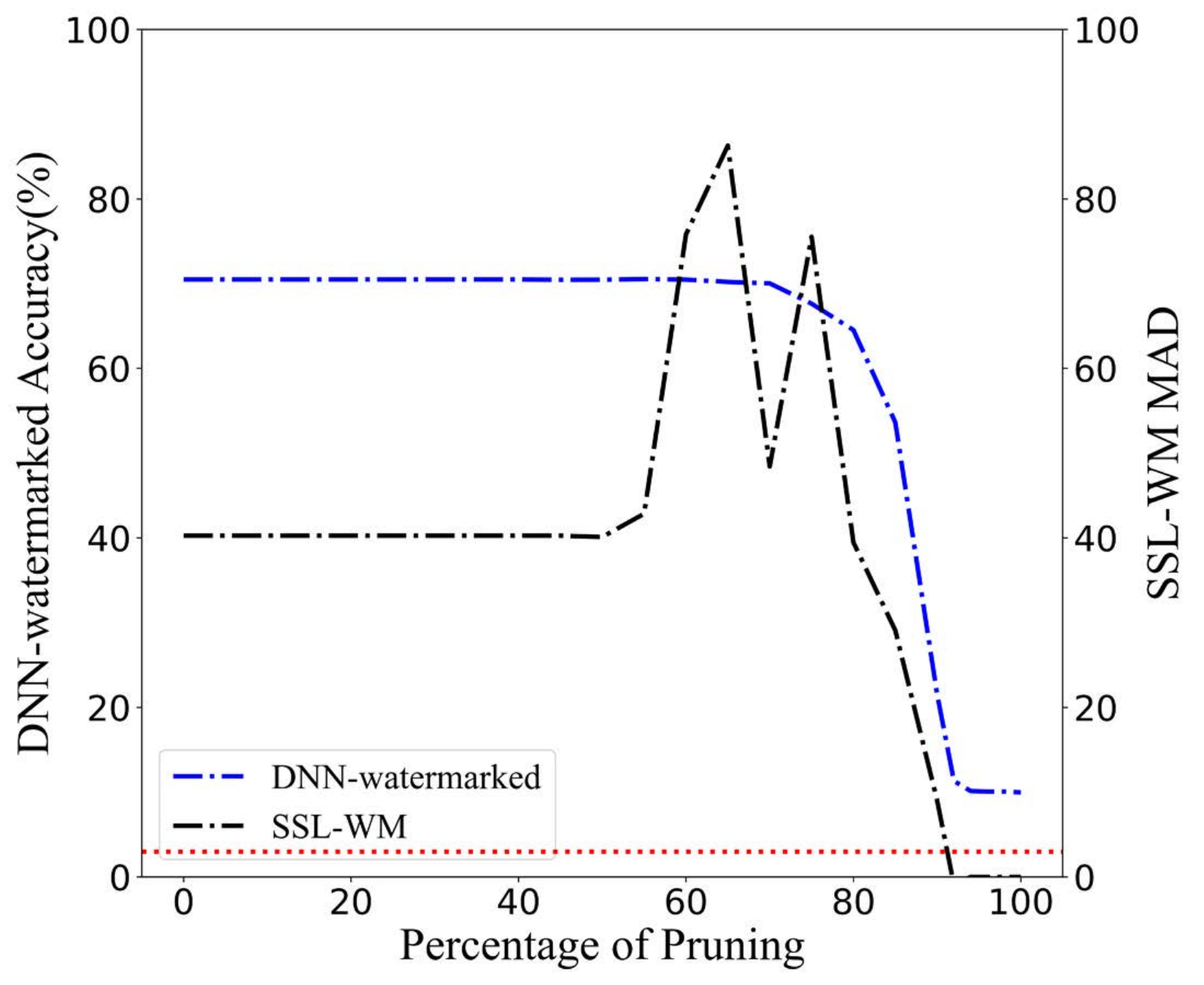}
\end{minipage}%
}%
\quad

\subfigure[Pruning IMDB]{
\begin{minipage}[t]{0.24\linewidth}
\centering
\includegraphics[width=1.8in]{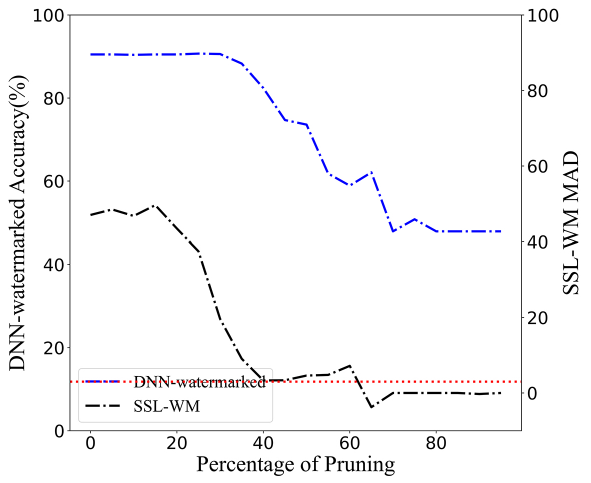}
\end{minipage}%
}%
\subfigure[Pruning SNLI]{
\begin{minipage}[t]{0.24\linewidth}
\centering
\includegraphics[width=1.8in]{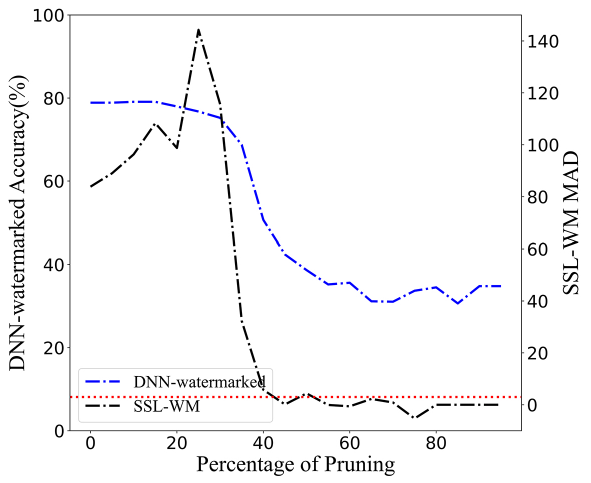}
\end{minipage}%
}%
\subfigure[Pruning MRPC]{
\begin{minipage}[t]{0.24\linewidth}
\centering
\includegraphics[width=1.8in]{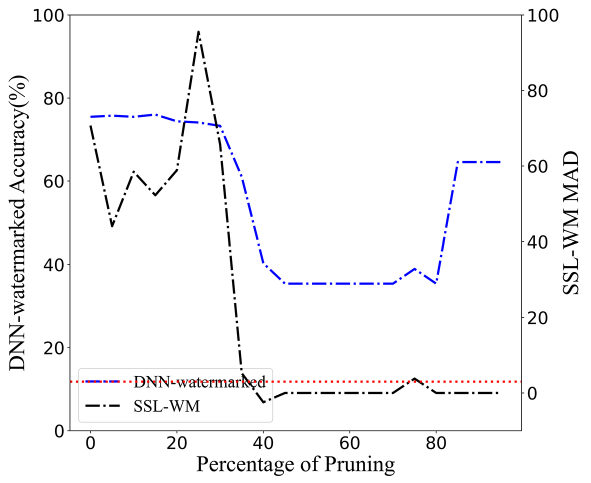}
\end{minipage}%
}%
\centering
\caption{Robustness against Fine-tuning ($\lambda=0.00001$) and Pruning. Red dotted line is the threshold of MAD to verify ownership.}
\label{fig:analysis_robustness}
\end{figure*}

\noindent\textbf{Performance of main tasks (Fidelity).}  
Ideally, SSL-WM should have a negligible impact on the main tasks' performance of the encoders-watermarked, i.e., a clean SSL model and a watermarked SSL model trained on the same dataset should have similar accuracy on the main task. 
The results in Table~\ref{tab:effectiveness} show that watermarked models perform similarly to the \ignore{lean}\add{clean} models on the main tasks, with average 0.41\% performance degradation. For SimCLR, MoCoV2, CLIP, and BERT, the performance of the watermarked models on the main tasks is slightly lower than that of the clean models, mainly because the watermark as an additional functionality affects the performance of the main task. However, the main task performance of the watermarked BYOL model is higher than that of the clean model, we suspect the reason as: BYOL updates the target network with a slow-moving average of the online network, and our watermarked loss introduces the random noise to the parameter changes of the target model, the random noise may cause the model to jump out of the local optimal solution and converge with better performance. And the performance of BiGAN on the downstream tasks is not as promising as the other contrastive-based models. In fact, in the BiGAN paper, the performance of BiGAN is also not as promising as the other contrastive-based models (BiGAN is only with 56.2\% accuracy on ImageNet, however, SimCLR is with 76.2\% accuracy on ImageNet).
Moreover, we analyze the reasons for the main task's good performance of the watermarked models. The feature representation and relationship between the watermarked samples and the background clean samples are not constrained during the training process.
The watermarked samples are mapped to a separate cluster as shown in Figure~\ref{analysis_simclr-tsne-downstream-tasks}. Thus, the watermarked SSL model has a similar representation capability as the clean SSL model on clean samples, and the main task performance of the watermarked model is as close as possible to the clean model. Moreover, though watermarked samples are mapped to a separate cluster, SSL-WM is still stealthy and cannot be effectively detected by watermark detection methods as shown in Section~\ref{subsec:stealthiness}.

\noindent\textbf{Time consumption (Efficiency).} 
The time consumption of watermarking is mainly reflected in two aspects: watermark embedding and ownership verification. 
On the one hand, the watermarking task will lead to additional time/computational consumption during the SSL model training process compared to the clean model training process. Table~\ref{tab:effectiveness} shows that the computational overhead introduced by the watermark is, on average, 1,991 minutes in value and 22.81\% in percentage, which are acceptable for the trainer since watermarking the model only needs to be done once.
\ignore{On the other hand, when ownership verification, if the watermark extraction process is too complex and time consuming, it may alert the attacker and thus lead to ownership verification failure.}\add{On the other hand, during ownership verification, if the watermark extraction process is overly complex and time-consuming, it may raise suspicion to attackers, probably resulting in a failure of ownership verification.} Therefore, we evaluate the watermark extraction time using our standard ownership verification process. As shown in Table~\ref{tab:effectiveness}, except CLIP and BERT, the extraction time of other SSL models is within 10 seconds. It takes a relatively long time (at most 57.72 seconds) for CLIP and BERT to extract watermarks because they are large models with numerous parameters to be compared with other models. In summary, the evaluation results demonstrate that our watermark extraction process is efficient\ignore{and beneficial for} during ownership validation.  
\subsection{Robustness}  
We assume attackers may be aware that the SSL models are watermarked. Thus they aim to remove the watermark by utilizing the watermark removing methods, such as fine-tuning and pruning.  Moreover, without revising the SSL models, attackers can perturb the inputs using input preprocessing attacks to evade the watermark verification. Particularly, the attackers usually have no access to the original training datasets of the SSL models, i.e., attackers will first transfer the SSL models to their downstream tasks and then use the downstream tasks' datasets to launch the removing attack methods. 



\begin{figure*}[h]
\centering

\subfigure[Watermark Pattern]{
\begin{minipage}[t]{0.16\linewidth}
\centering
\includegraphics[width=1.1in]{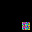}
\end{minipage}%
}%
\subfigure[NC-SimCLR]{
\begin{minipage}[t]{0.16\linewidth}
\centering
\includegraphics[width=1.1in]{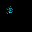}
\end{minipage}%
}%
\subfigure[NC-MoCoV2]{
\begin{minipage}[t]{0.16\linewidth}
\centering
\includegraphics[width=1.1in]{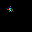}
\end{minipage}%
}%
\subfigure[NC-BYOL]{
\begin{minipage}[t]{0.16\linewidth}
\centering
\includegraphics[width=1.1in]{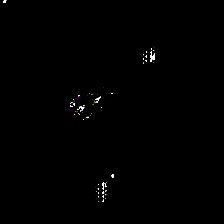}
\end{minipage}%
}%
\subfigure[ABS-SimCLR]{
\begin{minipage}[t]{0.16\linewidth}
\centering
\includegraphics[width=1.1in]{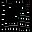}
\end{minipage}%
}%
\subfigure[ABS-MoCoV2]{
\begin{minipage}[t]{0.16\linewidth}
\centering
\includegraphics[width=1.1in]{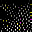}
\end{minipage}%
}%

\centering
\caption{Original watermark pattern and the reversed triggers by Neural Cleanse (NC) and ABS.}
\label{fig:trigger-pattern}
\end{figure*}

\noindent\textbf{Robustness against fine-tuning.} 
We assume that in the ideal scenario, attackers may have access to the testing dataset, e.g., test samples provided by the model owner to measure the model's performance, so they can leverage it to fine-tune the stolen model to destroy the watermark. Based on this, we randomly select 30\% of the test samples as the training dataset for fine-tuning and the rest of the test samples as the testing dataset. Then, we set the initial learning rate $\lambda=0.00001$ and the delay factor as 0.0005 to fine-tune watermarked DNNs using stochastic gradient descent (SGD). Finally, the downstream model tends to be stable during fine-tuning. The experimental results in Figure~\ref{fig:analysis_robustness} show that the main task accuracy on GTSRB gradually increases and finally stays stable because the SSL encoder is pre-trained on CIFAR-10. There is a significant data distribution difference between GTSRB and CIFAR-10. Fine-tuning all layers of the model using GTSRB will lead the model to fit better on GTSRB, but our SSL-WM is always far more extensive than 3, satisfying the ownership verification requirement. The main task accuracy on CIFAR-10, STL-10, CINIC-10, SNLI, and MRPC is stable during the fine-tuning attack, and our SSL-WM is far more extensive than 3 and can effectively verify its ownership. \add{We also use a larger learning rate $\lambda = 0.0001$ and re-run the experiment. The evaluation results are shown in Figure~\ref{fig:robustness4} of Appendix. Due to the relatively larger learning rate, the performance in the main task exhibits some instability during fine-tuning CIFAR-10, STL-10, and CINIC-10 when compared to $\lambda = 0.00001$. However, the MAD of our watermark still remains substantially larger than 3, satisfying ownership verification.}

\noindent\textbf{Robustness against pruning.} 
For pruning, we adopt the commonly used technique as in~\cite{han2015learning}, i.e., setting the parameters with smaller absolute values to zero, which has little impact on the model's performance. Taking the self-supervised algorithm SimCLR (in CV tasks) and BERT (in NLP tasks) as examples, we transfer them to their downstream tasks and prune 10\% to 90\% parameters of downstream tasks' models based on their absolute values. Results in Figure~\ref{fig:analysis_robustness} illustrate that if the watermarked DNNs are pruned at a lower pruning rate (less than 30\%), the performance of the main tasks is maintained. The outlier index of SSL-WM is far more extensive than 3 (i.e., the red line in Figure~\ref{fig:analysis_robustness}), which can effectively verify ownership of the stolen model. However, when the pruning rate is significant, the models have been pruned to be considered as ``fail'' on the main tasks (e.g., with 90\% pruning rate, the accuracy of CIFAR-10, STL-10, CINIC-10, GTSRB is reduced to 11.9\%, 22.3\%, 21.5\%, and 14.3\%, respectively; with 50\% pruning rate, the accuracy of binary classification NLP tasks (i.e., IMDB and MRPC) is reduced to below 60\% and the accuracy of the three classification NLP task (i.e., SNLI) is reduced to below 50\%), the outlier index of SSL-WM is less than 3. But such a high degree of pruning almost wholly ruins the main task of those models. We suspect the main reason for pruning robustness is that the neurons activated by watermarked samples are similar to those by the main tasks' samples. Thus SSL-WM will effectively verify ownership as long as the main tasks' performance of the model is maintained.

\noindent\textbf{Robustness against input preprocessing.}
Referring to~\cite{lukas2022sok}, we also evaluate the impact of input preprocessing on our watermarking approach, e.g., adding Gaussian noise to the entire input image. However, Gaussian noise will cause blur in the input images, thus downgrading the accuracy of the main task. We consider that the maximum acceptable performance degradation of the main task for the attacker is 3\%, so we adjust the standard deviation of Gaussian noise (the mean is 0) to control the degradation within 3\%. After launching this attack, the accuracy of the watermarked models based on SimCLR for CIFAR-10, STL-10, GTSRB, and CINIC-10 tasks is 83.00\%, 71.61\%, 61.56\%, and 70.19\% respectively, and MAD is 167.50, 22.97, 54.15 and 41.10, respectively, satisfying ownership verification.

\subsection{Stealthiness}
\label{subsec:stealthiness}

After stealing target models, attackers may detect whether the watermark exists in models \add{either at the model level or the input level}. Since our watermark pattern is similar to the trigger pattern in backdoor attacks against DNNs, at the model level, attackers may directly detect the presence of our watermark using backdoor detection approaches, such as Neural Cleanse~\cite{wang2019neural}, ABS~\cite{liu2019abs} \add{and MNTD~\cite{xu2021detecting}}. \add{At the input level,} the attackers can use \add{Februus~\cite{doan2020februus}, Beatrix~\cite{ma2022beatrix}}, and Anomaly Detection~\cite{jia2021entangled} to detect watermarked inputs, thus evading ownership verification. We evaluate the stealthiness of SSL-WM against them as below. 


\noindent\textbf{Neural Cleanse.}
Neural Cleanse reconstructs potential triggers by reverse engineering and then uses them to identify and remove backdoors.
We apply Neural Cleanse to our watermarked SimCLR, MoCoV2, BYOL, and BiGAN on the CIFAR-10 downstream task to generate the trigger patterns for the suspect labels and use the Mask Jaccard Similarity (MJS)~\cite{pang2020trojanzoo} to measure the intersection between the reversed trigger patterns and the actual watermark pattern. Note that Neural Cleanse requires the clean samples related to the main task to reconstruct triggers, so we use the test dataset of CIFAR-10 as its clean dataset in our evaluation. Figure~\ref{fig:trigger-pattern} shows the reversed triggers and our watermarked pattern, and Table~\ref{tab:backdoor-detection} illustrates the MJS of these reversed triggers. The experimental results \ignore{prove}\add{show} that Neural Cleanse does not find any suspect labels in BiGAN but finds suspect labels and generates the triggers for SimCLR, MoCoV2, and BYOL. However, these reversed triggers are not similar to the watermark pattern, and the largest MJS is only 0, which means that there is no intersection between the reversed trigger and the original watermark pattern. Then, we also utilize the generated triggers to remove the watermark using the two approaches proposed in Neural Cleanse: adversarial training-based approach and pruning-based approach. When launching adversarial training of Neural Cleanse, the outlier index of these models is still larger than 3 (i.e., 19.31 on SimCLR, 35.67 on MoCoV2, and 9.877 on BYOL). When launching pruning of Neural CLeanse, the outlier index of these models is also larger than 3 (i.e., 38.16 on SimCLR, 29.58 on MoCoV2, and 4.52 on BYOL). Thus, Neural Cleanse cannot effectively generate high-fidelity triggers to detect our watermark. We argue that the main reason is that our watermarked samples are trained to be represented as a separate feature representation cluster in the watermarked encoders. These watermarked samples do not belong to a high-confidence label, so the trigger generated for the target high-confidence label (i.e., maximum activation) tends to differ from the watermark pattern.
 


\begin{table}[t]
\begin{threeparttable}
\centering
\footnotesize
\caption{Backdoor Detection}
\label{tab:backdoor-detection}
\begin{tabular}{m{2cm}
<{\centering}|m{2cm}
<{\centering}|m{3cm}
<{\centering}}
\hline
\textbf{SSL Models}&\textbf{Detection Methods}& \textbf{Mask Jaccard Similarity}\\
\hline \hline
\multirow{2}{*}{SimCLR}&\textbf{Neural Cleanse}& \text{0} \\ \cline{2-3}
\textbf{}& \textbf{ABS}& \text{0.090}   \\ \hline
\multirow{2}{*}{MoCoV2}&\textbf{Neural Cleanse}& \text{0} \\ \cline{2-3}
\textbf{}& \textbf{ABS}& \text{0.065}   \\ \hline
\multirow{2}{*}{BYOL}&\textbf{Neural Cleanse}& \text{0} \\ \cline{2-3}
\textbf{}& \textbf{ABS}& \text{-} \\ \hline
\multirow{2}{*}{BiGAN}&\textbf{Neural Cleanse}& \text{-} \\ \cline{2-3}
\textbf{}& \textbf{ABS}& \text{-} \\ \hline
\end{tabular}
\end{threeparttable}
\begin{tablenotes}
\footnotesize
\item[]{``-'' represents that the backdoor detection regards the model as clean and does not generate any trigger.}

\end{tablenotes}
\end{table}

\noindent\textbf{ABS.}
ABS detects whether a suspected DNN model has a backdoor by analyzing the behavior of internal neurons.
We use ABS to detect and generate trigger patterns against our watermarked models on CIFAR-10. We utilize the test dataset of CIFAR-10 as the background for ABS to generate triggers in our evaluation. Figure~\ref{fig:trigger-pattern} shows the generated triggers of ABS and Table~\ref{tab:backdoor-detection} illustrates the MJS of these reversed triggers. ABS does not find suspect labels in both BYOL and BiGAN, but only finds suspect labels and generates the triggers for SimCLR and MoCoV2. 
\ignore{But there is too much noise in these generated triggers, and the MJS of these triggers is only 0.078 on average. Moreover, the triggers generated by ABS have worse fidelity than those generated by Neural Cleanse.}\add{The generated triggers are quite different from our true watermark pattern as shown in Figure~\ref{fig:trigger-pattern}, and the MJS of these triggers is only 0.078 on average\footnote{Smaller value indicates there is very little similarity between the generated trigger and the real watermark pattern.}. As admitted in the ABS paper, it occasionally reverse-engineers (strong) benign features and considers them as a trigger. We believe these triggers are actually derived from such benign features, making ABS ineffective in detecting our SSL-WM.} Actually, the inefficacy of ABS is mainly because our watermark is embedded by influencing numerous neurons of target models; while the trigger pattern generated by ABS is based on one or some compromised neurons only. Thus, it is difficult for ABS to generate high-fidelity triggers similar to our watermarks.


\noindent\add{\textbf{MNTD.} MNTD~\cite{xu2021detecting} aims to train a meta-classifier that takes the representation of the target model as the input and performs a binary classification to determine if the target model is backdoored. We evaluate MNTD by examining if it can detect a watermarked (i.e., backdoored) encoder or not. Following the default experimental settings of MNTD, we use the meta-classifier released by MNTD to detect our watermarked encoders trained by SimCLR, MoCoV2, BYOL and BiGAN. These watermarked models are with the ResNet-18 architecture, the same as the one used in MNTD paper. According to our evaluation, MNTD cannot effectively detect our watermark from the watermarked encoders, i.e., 0\% detection accuracy. It is noteworthy that the training of the meta-classifier is based on feature representation from backdoors obtained through supervised models, as in the released code. However, our watermark feature representation is from the encoders trained by self-supervised learning. The feature difference between supervised learning and self-supervised learning could lead to the failure of MNTD's watermark detection.}

\noindent\add{\textbf{Beatrix.} Beatrix~\cite{ma2022beatrix} aims to detect backdoor by Gram matrix, which can identify the poisoned samples by capturing the anomalies in activation features, so the attackers can utilize Beatrix to detect the watermarked inputs. Following the default experimental settings of Beatrix on CIFAR-10, we evaluate Beatrix against our watermarked encoders with the ResNet18 architecture on the CIFAR-10 task. The evaluation results indicate that Beatrix cannot effectively detect watermarked samples, i.e., the true positive rate is only 0.2\%, 0.2\%, 0\%, and 0.4\% for encoders trained by SimCLR, Moco V2, BYOL, and BiGAN, respectively. We discussed the results with the authors of Beatrix, and they suggested that the detection failure might be due to a significant entanglement of the features between our watermarks and the main task in the last layer of encoders.}


\noindent\add{\textbf{Februus.} Februus~\cite{doan2020februus} sanitizes the incoming inputs by removing the potential trigger identified by Grad-CAM, which locates the critical regions that contribute significantly to the classification results. Then, it obtains the clean inputs by reconstructing the removed region of the inputs using GAN. 
We embed our watermark into encoders with the ResNet18 architecture, trained using SimCLR, MoCo V2, BYOL, and BiGAN. Then, we evaluate Februus using its released code~\cite{code-februus} against our watermarked encoders in the CIFAR-10 task. Before using Februus, the accuracy of the watermarked models is 83.24\%, 83.43\%, 86.56\%, and 51.27\%, respectively, and the MAD of the watermark is 76.60, 24.92, 17.48, and 9.73, respectively. We launch Februus using the default settings, and the clean data accuracy of the watermarked model drops to 74.93\%, 74.24\%, 76.56\%, and 48.82\%, respectively. We believe this is because Grad-CAM cannot always correctly locate the regions of the watermark pattern, and GAN cannot always generate a reconstructed image that is similar to the clean sample. Meanwhile, the MAD of our watermark is 41.77, 11.73, 8.27, and 4.66, respectively, still verifying ownership successfully. Thus, Februus cannot effectively evade our watermark verification.}

\begin{table}[!h]
\begin{threeparttable}
\centering
\footnotesize
\caption{Watermarked Inputs Detection by LOF}
\label{tab:outlier-detection}
\begin{tabular}{m{1.5cm}
<{\centering}|m{1.7cm}
<{\centering}|m{1.7cm}
<{\centering}|m{1.7cm}
<{\centering}}
\hline
\textbf{SSL Models}& \textbf{False Positive}& \textbf{Watermark Detection}&\textbf{Accuracy Loss}\\
\hline \hline

\textbf{SimCLR}& \text{0$\%$} &\text{0$\%$}  &\text{0.00\%}\\ \hline
\textbf{MoCoV2}& \text{3.25$\%$} &\text{3.50$\%$} & \text{3.00\%}  \\ \hline
\textbf{BYOL}& \text{1.75$\%$} &\text{1.00$\%$}& \text{0.00\%} \\ \hline
\textbf{BiGAN}& \text{0.75$\%$} &\text{1.00$\%$}&\text{0.25\%}  \\ \hline
\end{tabular}
\end{threeparttable}
\begin{tablenotes}
\footnotesize

\item[]{Watermarked Inputs Detection represents the probability that the watermarked input samples are detected correctly. False Positive represents the probability that the clean input samples are detected as the watermarked input samples. Accuracy Loss represents the performance degradation of the models. }

\end{tablenotes}
\vspace{5pt}
\end{table}

\noindent\textbf{Anomaly Detection.} 
Referring to~\cite{jia2021entangled}, we use the commonly used outlier detectors Local Outlier Factor (LOF)~\cite{breunig2000lof} to detect the abnormal inputs of watermarked models on the CIFAR-10 downstream task, using the activation of the last linear layer. We show the results in Table~\ref{tab:outlier-detection}. In general, the average detection rate of the watermark is only 1.38\%, and this leads to a small degradation of the main task performance by an average of 0.81\%. Thus, we think LOF cannot effectively detect our watermarked inputs. We suspect this is because watermarked samples are with similar feature vectors, and the density around the feature vector of each watermarked sample is similar, thus LOF will not view the watermarked samples as outliers.


\subsection{Comparison with State-of-the-art}
\label{subsec:comparison}

We compare our SSL-WM with the state-of-the-art, i.e., SSLGuard. In particular, we train watermarked ResNet18 encoders using the released source code of SSLGuard and our SSL-WM, respectively. Then we use our watermark verification approach\footnote{SSLGuard does not have a watermark verification approach in downstream tasks, since it only detects watermark from encoders directly.} to examine the embedded watermark in different downstream tasks, including CIFAR-10, CINIC-10, GTSRB, and STL-10. The watermark pattern occupies 1.56\% of the input samples. For the watermarked encoders trained by SSLGuard, we calculate the outlier index according to Equation~(\ref{loss:mad}), and obtain -0.675, -0.674, 0.305, and 3.512 for CIFAR-10, CINIC-10, GTSRB, and STL-10, respectively. The outlier index values of CIFAR-10, CINIC-10, and GTSRB are smaller than 3, thus failing to verify the ownership of watermarked encoders. In contrast, for the watermarked encoders trained by our SSL-WM, the outlier index values are 88.95, 47.06, 68.01, and 50.17 for CIFAR-10, CINIC-10, GTSRB, STL-10, respectively, all larger than 3, satisfying ownership verification. Thus, SSL-WM outperforms SSLGuard in verifying ownership of encoders reused in downstream tasks.

\subsection{Watermark Embedding in Feature Space}
\label{subsec:watermark-embedding-feature-space}

To show our watermark embedding performance of SSL models, we use T-SNE to show the representations space of the clean and watermarked samples for each watermarked encoder (including SimCLR, MoCoV2, BYOL, and BiGAN) on CIFAR-10 in Figure~\ref{fig:watermarked-four-ssl-models} of Appendix.  Figure~\ref{fig:watermarked-four-ssl-models} shows that the watermarked samples are a separate cluster in the embedding space of all these watermarked encoders, compared with the clean models, indicating that our SSL-WM has mapped the watermarked samples to the invariant embedding space of the watermarked encoder. Though watermarked BYOL encoder projects a few watermarked inputs next to clean examples, it projects major watermark examples far away from clean ones and to a separate cluster, which is enough to distinguish it from the benign encoder.
Particularly, SimCLR, MoCoV2, and BYOL are contrastive-based, while BiGAN is generative-based, demonstrating that our SSL-WM can be broadly applied to protect the IP of both kinds of SSL algorithms.

\section{Discussion}

\noindent\textbf{Defense against ambiguity attack.}  
\add{
The adversary can generate a trigger and use it to forge a valid SSL-WM-style watermark. If this forged watermark demonstrates high accuracy, the adversary can use it to falsely claim ownership or dispute ownership claims from the true owner. We evaluate this ambiguity attack by generating a Universal Adversarial Patch (UAP) using optimization methods like gradient descent and using it as a forged watermark against~\cite{wu2022watermarking}. In particular, \cite{wu2022watermarking} inserts an untargeted backdoor into encoders as the watermark. This backdoor causes downstream classifiers to classify the watermarked inputs (i.e., inputs with the trigger attached) as any label other than the ground truth. According to our evaluation, against the watermarked encoder with the ResNet18 architecture, the forged watermark (i.e., the UAP) achieves a watermark accuracy of 92.32\% in downstream tasks, even higher than the average watermark accuracy (i.e., 89.01\%) of the true watermark pattern used in~\cite{wu2022watermarking}. Therefore, such an ambiguity attack indeed tampers with the ownership verification from the true owner, so the defense against it is considered in various watermarking approaches~\cite{fan2019rethinking,ong2021protecting,lv2023robustness}.}

To defeat the ambiguity attack, we can constrain the generation of watermark patterns. For instance, a hash function can be used to generate the watermark trigger pattern $wm = Hash(info, key)$, where $info$ is related to the owner's identity (e.g., name) and $key$ is the secret key only known to the owner. The watermark pattern generated by such a hash function makes it challenging to forge SSL-WM.
On the one hand, even if the adversary can train a UAP against the target feature vector of the stolen encoder and use it as a trigger, it is computationally hard for the adversary to obtain $info$ and $key$ by reversing the hash of the UAP trigger to claim ownership, especially $info$ should have semantics.
On the other hand, the adversary can generate a semantic $info$ and randomly choose a $key$ to produce a forged trigger pattern. However, such a forged trigger pattern cannot be well recognized by the watermarked SSL encoder since it is not in the training dataset or a UAP. \ignore{Thus the output distribution of the samples stamped with this trigger will not have a small dispersion or failure to verify ownership.} \add{Consequently, the output distribution of the samples attached by this trigger will not have a small dispersion, so it cannot be used to claim ownership fraudulently.} \add{Note that the above method can be adopted by other watermarks, like~\cite{wu2022watermarking} to improve their resilience against the ambiguity attack.}

Alternatively, the attacker can also select a watermark pattern and embed the pattern into the stolen DNN following our watermark injection algorithm. In the case of disputed ownership, the attacker can use this forged watermark to claim ownership of the stolen model.
\ignore{However, due to the high stealthiness of our watermark pattern (see Section), it is unlikely that the adversaries' forged watermarked pattern can overwrite the owner's watermarked pattern.} 
\add{However, our watermark pattern is highly stealthy, as Neural Cleanse and ABS cannot reconstruct a trigger similar to our true watermark pattern as shown in Section~\ref{subsec:stealthiness}. Thus, the attacker cannot inject a forged watermark while removing our true watermark, because the forged watermark patterns (either the reversed trigger or the watermark pattern selected by the adversary) are not similar to our true watermark patterns at all. Thus, the forged watermark embedded by the attacker cannot overwrite our true watermark since their patterns are totally different.} 
In this case, the attacker can only present the DNN containing his/her watermark and our watermark, while we can present the DNN containing our watermark only. Thus, it is clear who is the true owner.

\noindent\add{\textbf{Resilience to Confidence Perturbation Attack.} Aware of our watermarking approach, a naive approach for the attackers would be to randomize the classification result for each watermarked input to evade the watermark verification. However, as evaluated in Section V-D, existing approaches cannot detect our watermarked samples, so the only option for the attackers is to randomize the classification results for all queries (including clean queries without our watermark). As in Equation~(\ref{loss:shannon-entropy}), we employ hard labels instead of soft labels (i.e., output confidence) for ownership verification. This means attackers must introduce substantial noise into the output confidence values to alter the output labels, thereby invalidating any potential watermarked queries. However, this may significantly impact the model's primary task performance. We evaluate such an attack, and the evaluation results show that when Gaussian noise is added to the output confidence, the accuracy of the model drops dramatically across multiple datasets, i.e., 46.22\% on CIFAR-10, 38.35\% on CINIC-10, 34.21\% on STL-10 and 13.74\% on GTSRB, while our watermark still effectively verifies ownership with MAD values of 14.19 on CIFAR-10, 7.81 on CINIC-10, 18.75 on STL-10, and 26.18 on GTSRB.}

\section{~~Conclusion}
\label{sec:Conclusion}

In this paper, we propose a novel black-box watermark approach SSL-WM, to protect the ownership of encoders pre-trained by self-supervised learning, in an unknown and diverse downstream tasks scenario. SSL-WM trains SSL encoders to map the watermarked samples to an invariant watermark representation space and verifies the ownership of the suspect models by querying them with the watermarked samples in a black-box manner. Moreover, we evaluate our watermark on both CV and NLP tasks, using various self-supervised learning algorithms, including contrastive-based, and generative-based algorithms. The results demonstrate that our watermark can effectively verify the ownership of encoders trained by self-supervised learning algorithms, and no existing watermarks achieve similar results as ours.



\section*{Acknowledgment}

We thank the Shepherd and reviewers for their constructive feedback. The IIE authors are supported in part by NSFC (92270204, 62302497), Beijing Natural Science Foundation (No.M22004), Youth Innovation Promotion Association CAS, a research grant from Huawei, and Beijing Nova Program.




%
\bibliographystyle{IEEEtranS}
\bibliography{reference.bib}



\section*{Appendix}
\label{sec:Appendix}




\begin{figure*}[ht]
\centering

\subfigure[Fine-tuning CIFAR-10]{
\begin{minipage}[t]{0.24\linewidth}
\centering
\includegraphics[width=1.8in]{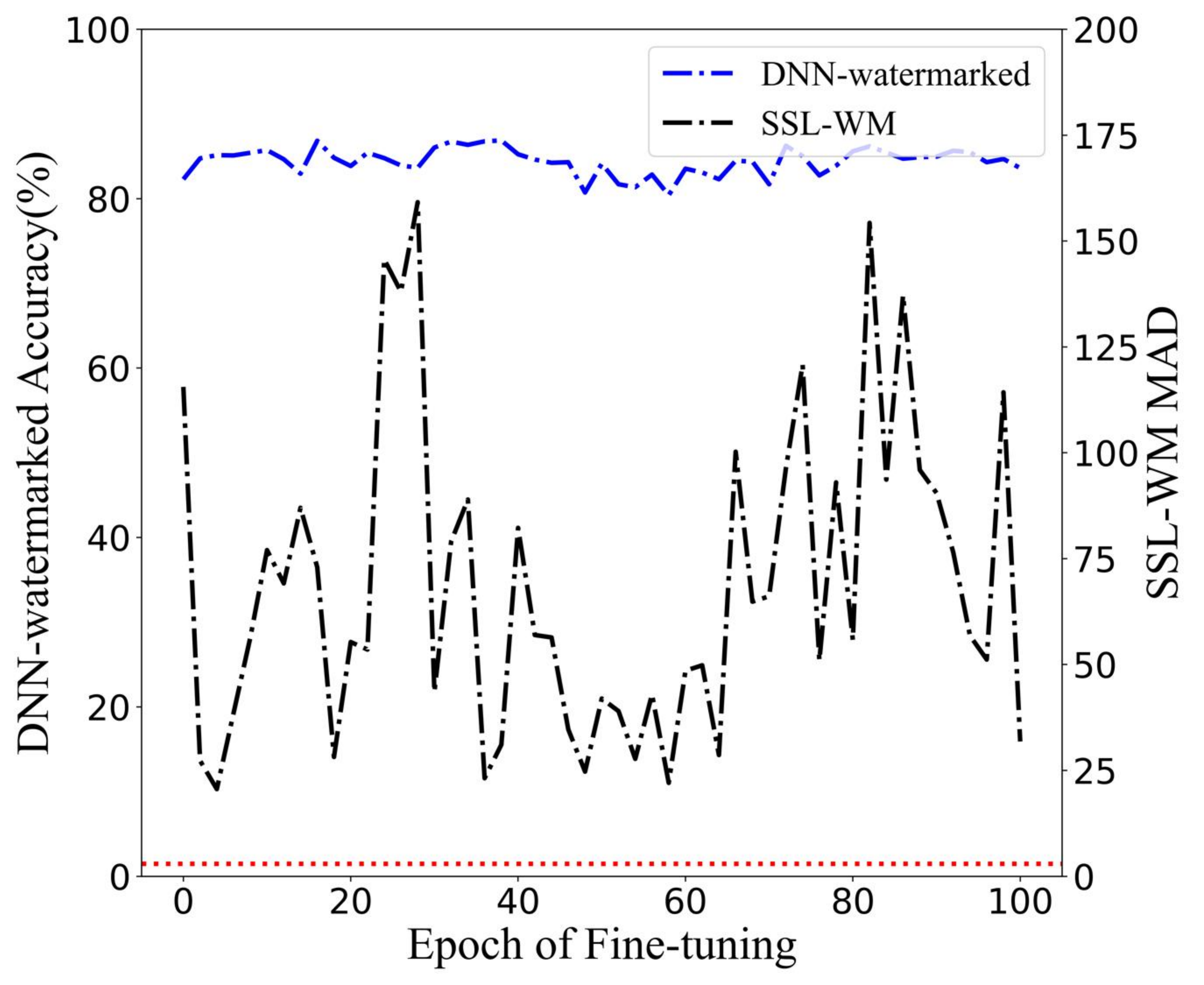}
\end{minipage}%
}%
\subfigure[Fine-tuning STL-10]{
\begin{minipage}[t]{0.24\linewidth}
\centering
\includegraphics[width=1.8in]{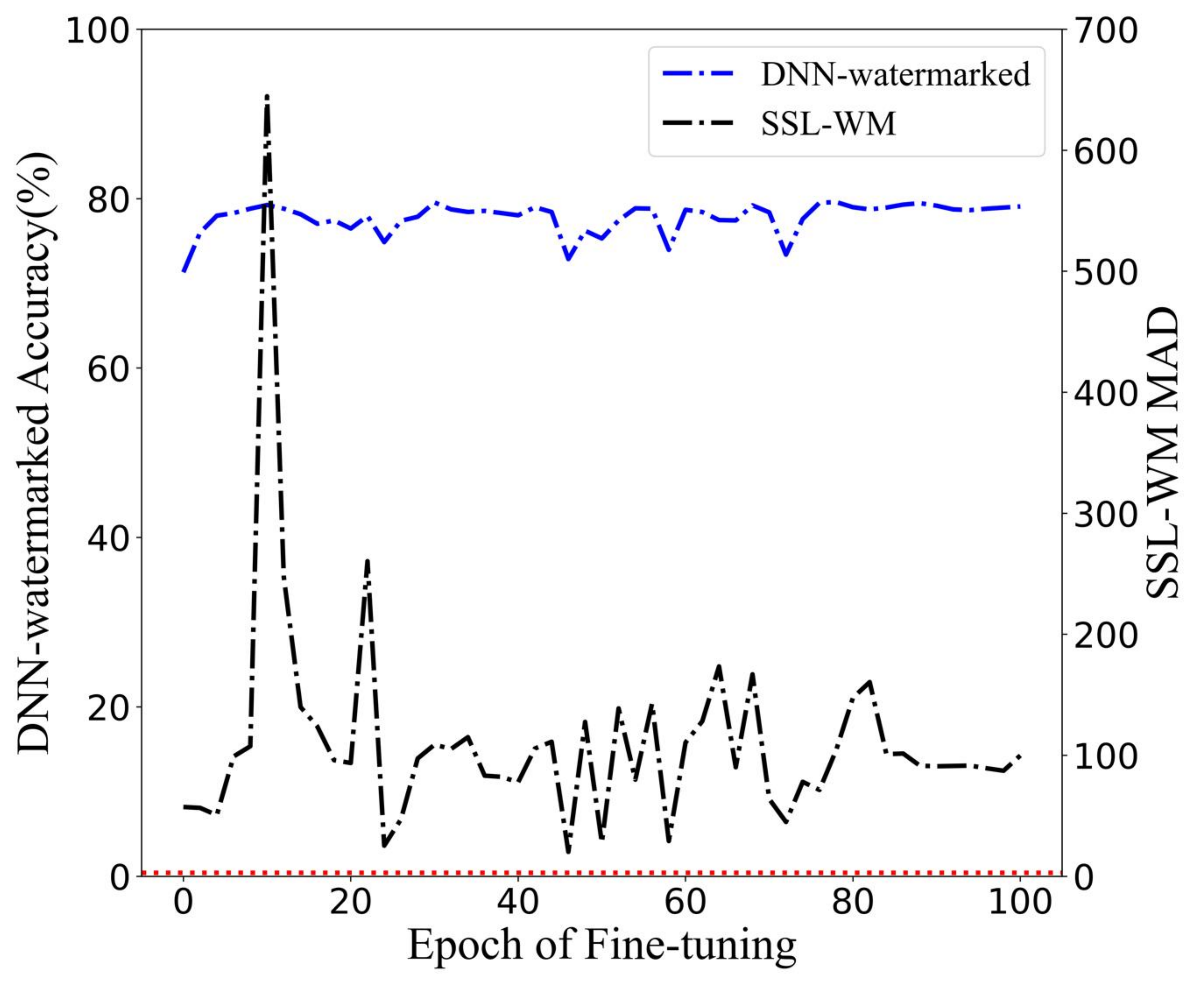}
\end{minipage}%
}%
\subfigure[Fine-tuning GTSRB]{
\begin{minipage}[t]{0.24\linewidth}
\centering
\includegraphics[width=1.8in]{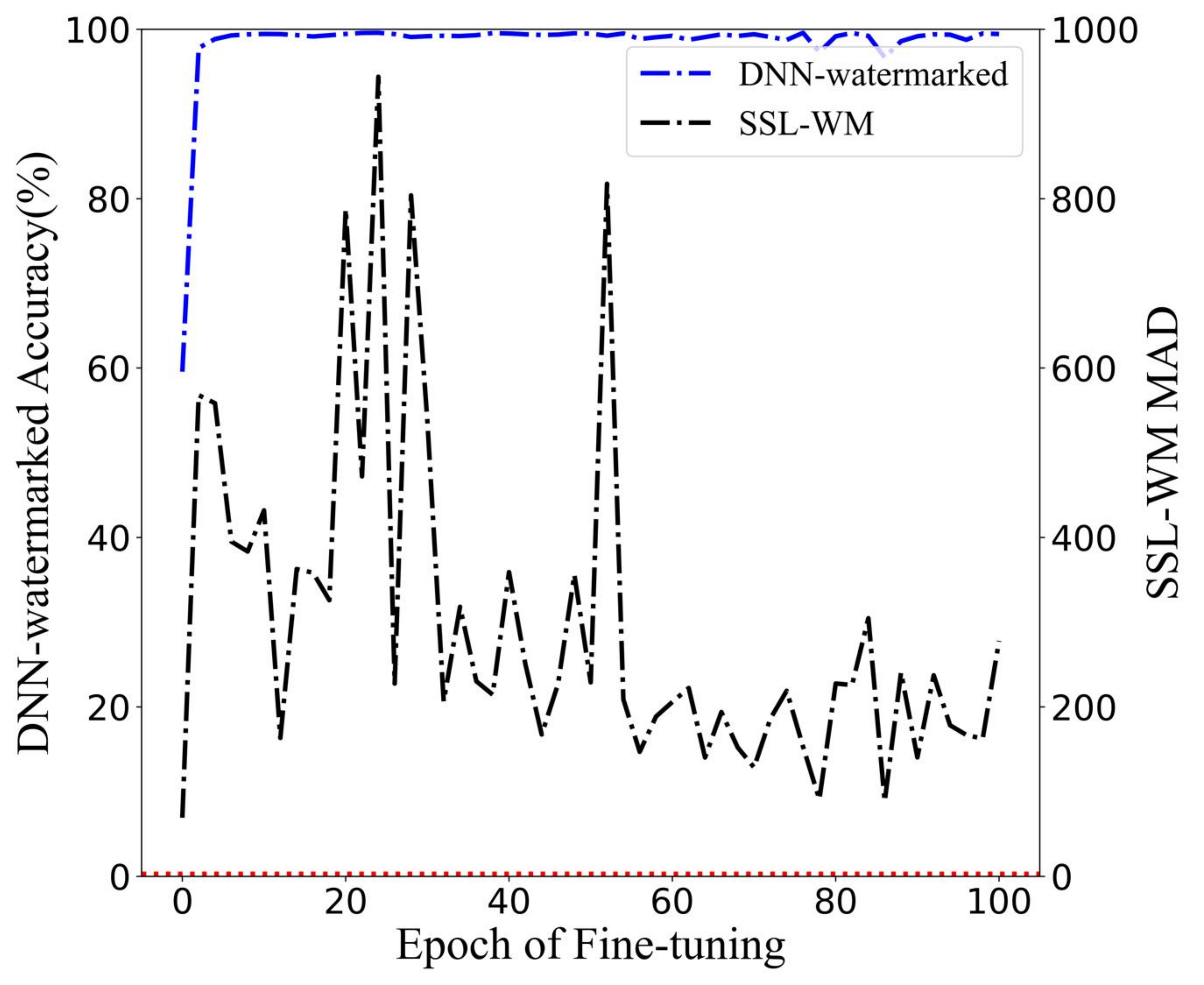}
\end{minipage}%
}%
\subfigure[Fine-tuning CINIC-10]{
\begin{minipage}[t]{0.24\linewidth}
\centering
\includegraphics[width=1.8in]{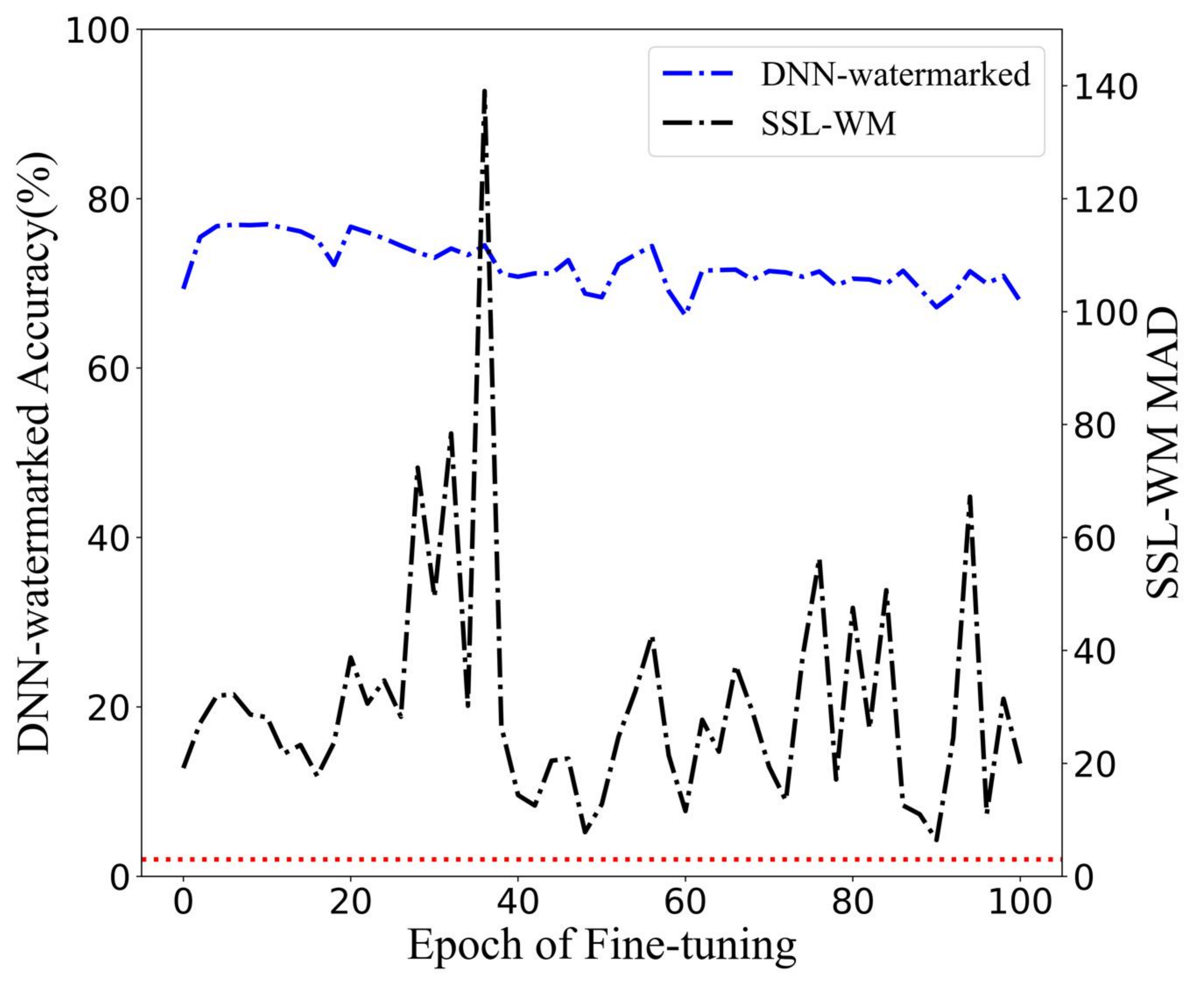}
\end{minipage}%
}%
\quad

\subfigure[Fine-tuning IMDB]{
\begin{minipage}[t]{0.24\linewidth}
\centering
\includegraphics[width=1.8in]{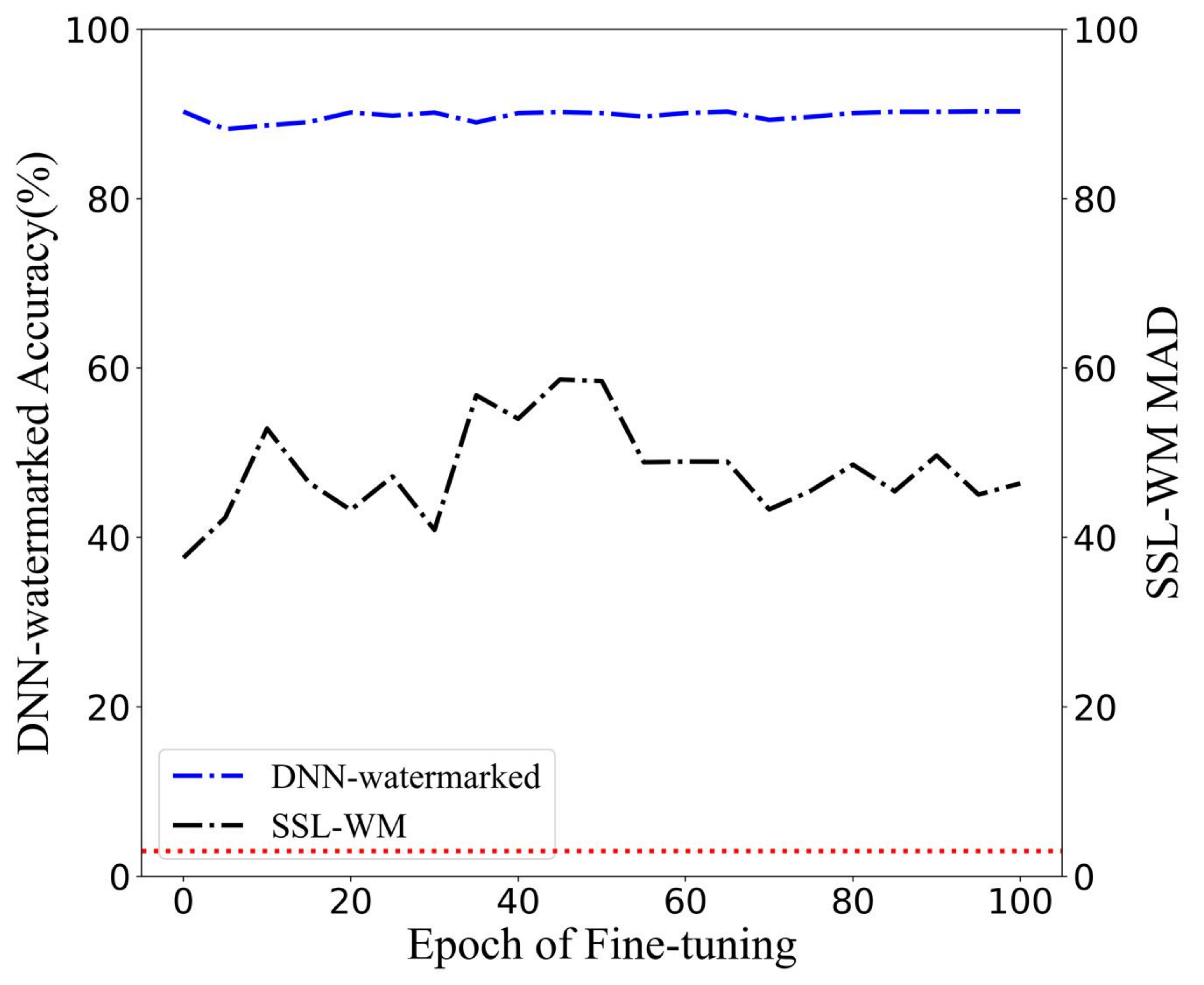}
\end{minipage}%
}%
\subfigure[Fine-tuning SNLI]{
\begin{minipage}[t]{0.24\linewidth}
\centering
\includegraphics[width=1.8in]{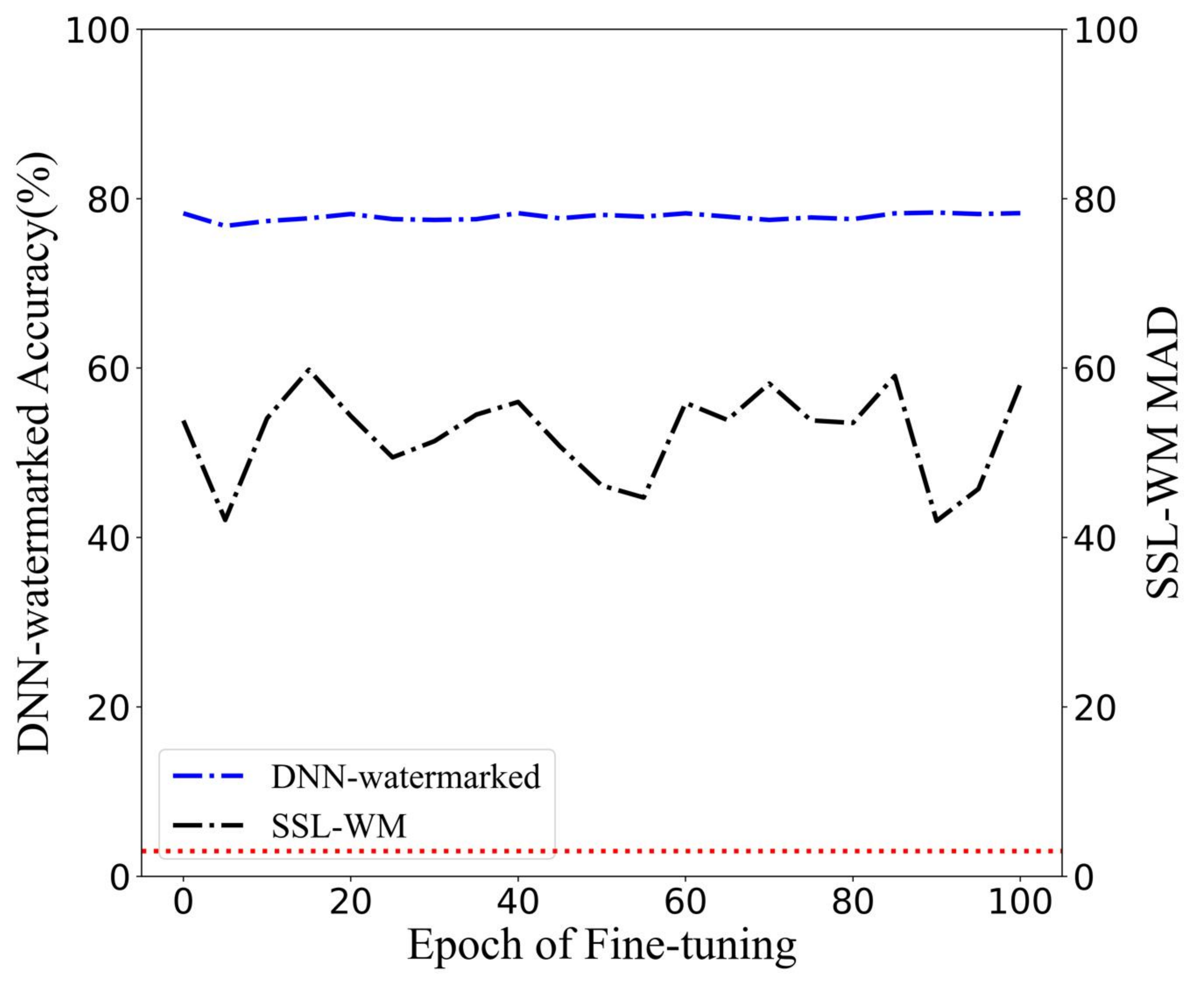}
\end{minipage}%
}%
\subfigure[Fine-tuning MRPC]{
\begin{minipage}[t]{0.24\linewidth}
\centering
\includegraphics[width=1.8in]{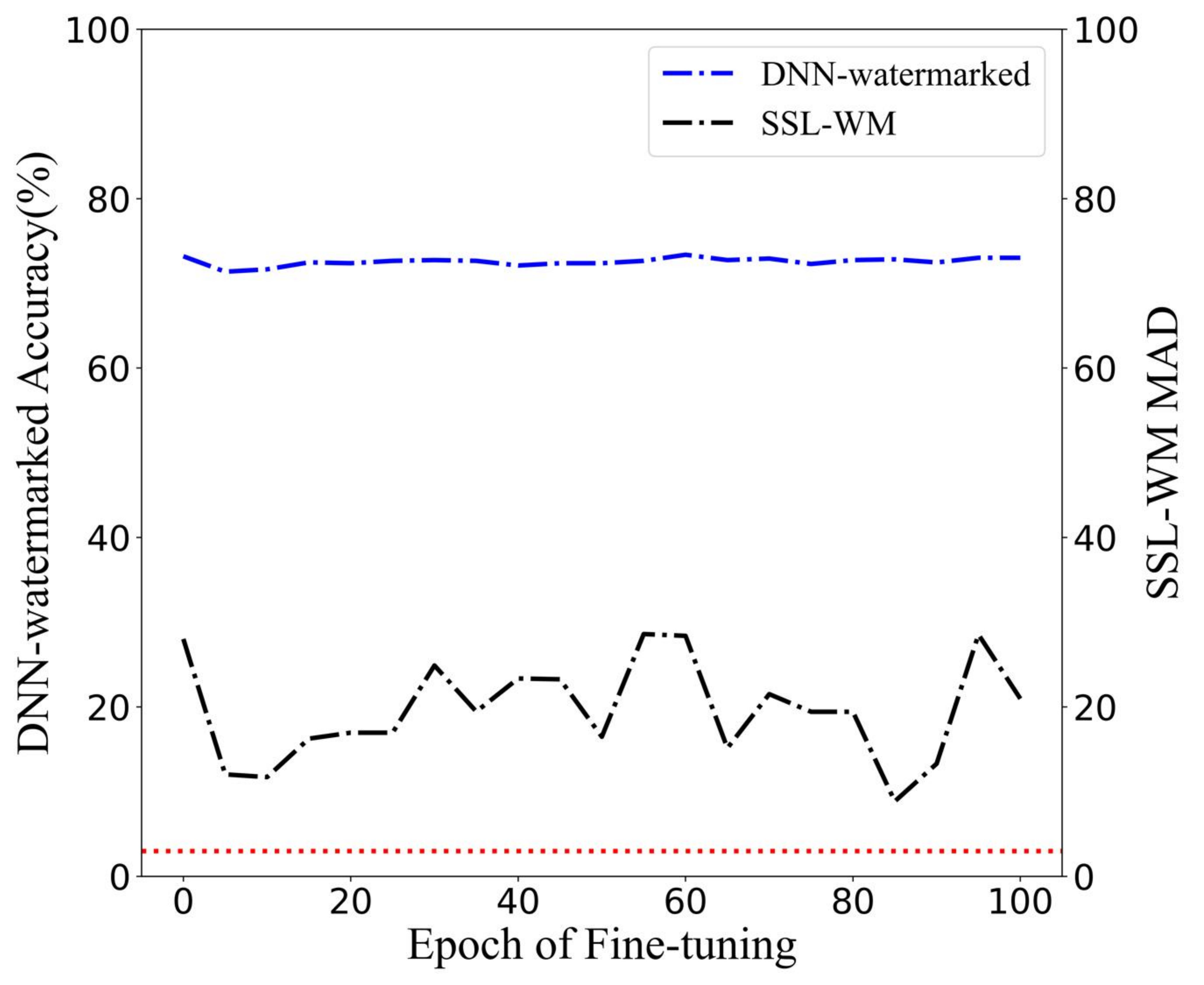}
\end{minipage}%
}%

\centering
\caption{Robustness against Fine-tuning ($\lambda=0.0001$). The red dotted line is the threshold of MAD to verify ownership.}
\label{fig:robustness4}
\vspace{-26pt}
\end{figure*}


\begin{figure*}[htb]
\centering

\subfigure[SimCLR-Clean]{
\begin{minipage}[t]{0.25\linewidth}
\centering
\includegraphics[width=1.8in]{fig/simclr_TSNE_cifar10_simclr-encoder-clean-891.pth.pdf}
\end{minipage}%
}%
\subfigure[MoCoV2-Clean]{
\begin{minipage}[t]{0.25\linewidth}
\centering
\includegraphics[width=1.8in]{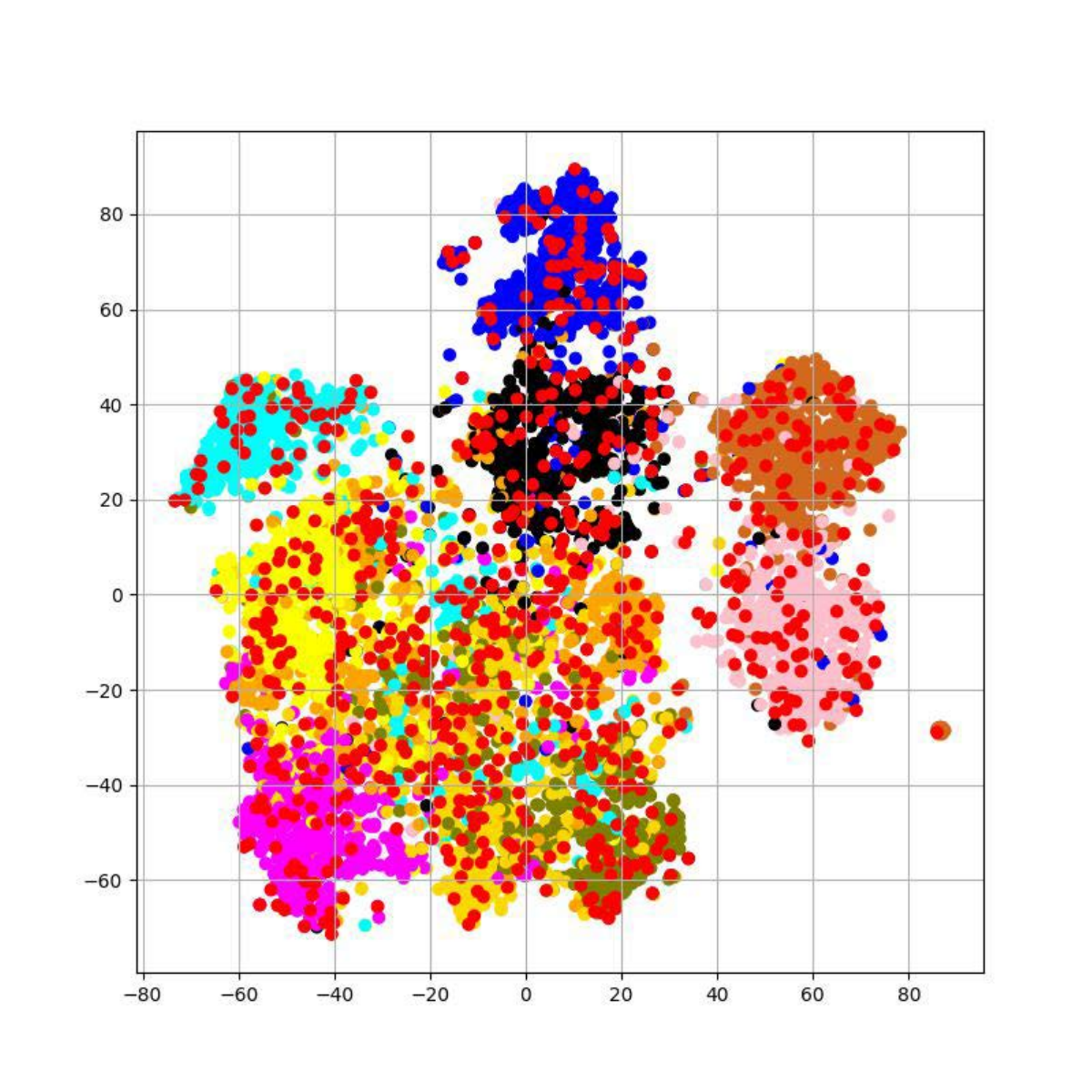}
\end{minipage}%
}%
\subfigure[BYOL-Clean]{
\begin{minipage}[t]{0.25\linewidth}
\centering
\includegraphics[width=1.8in]{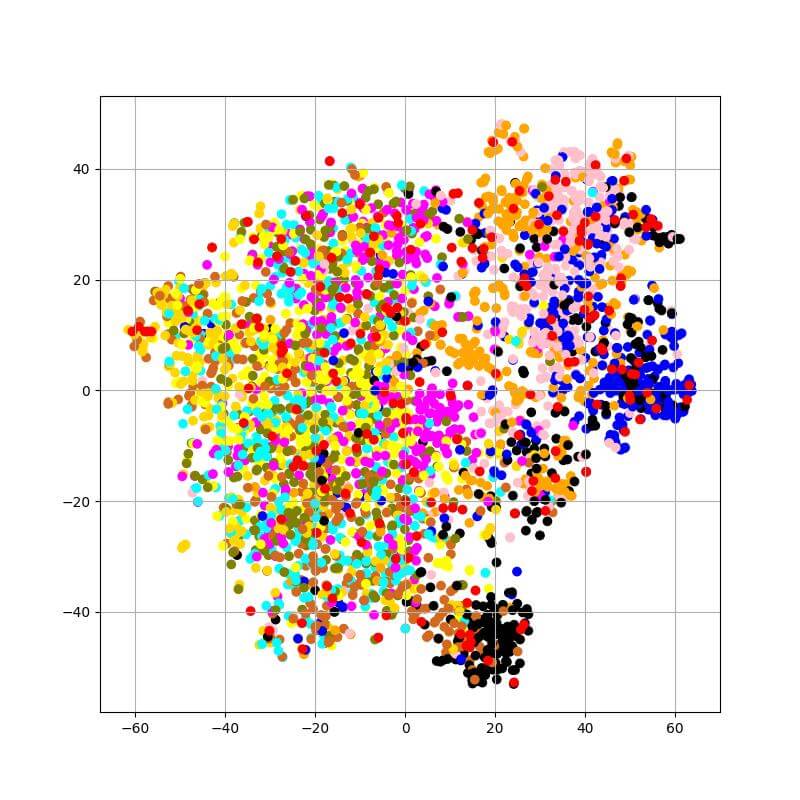}
\end{minipage}%
}%
\subfigure[BiGAN-Clean]{
\begin{minipage}[t]{0.25\linewidth}
\centering
\includegraphics[width=1.8in]{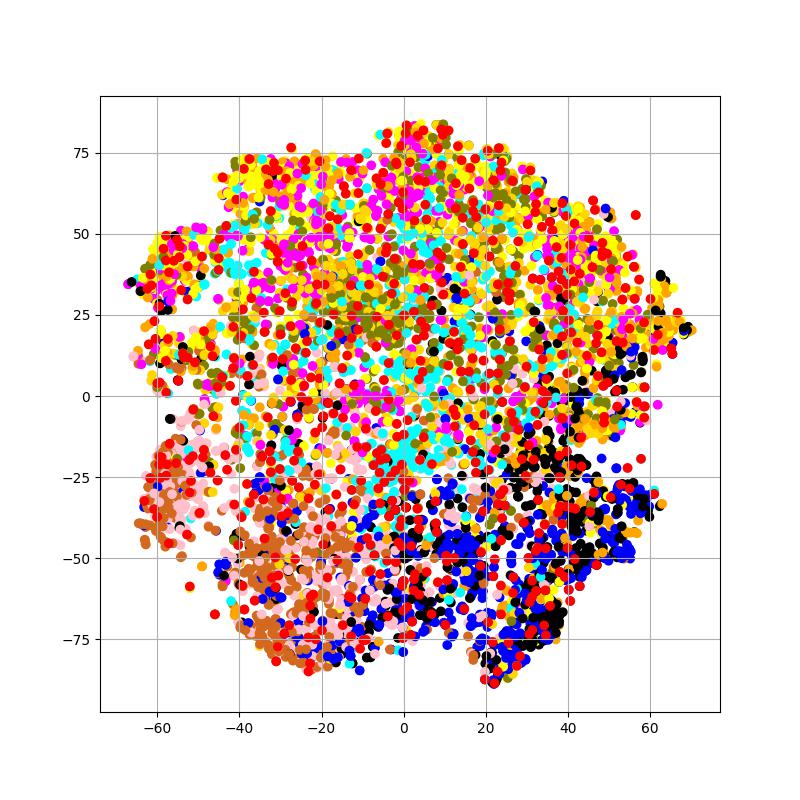}
\end{minipage}%
}%

\quad
\vspace{-10pt}

\subfigure[SimCLR-Watermarked]{
\begin{minipage}[t]{0.25\linewidth}
\centering
\includegraphics[width=1.8in]{fig/simclr_TSNE_cifar10_simclr-encoder-391-v3.pth.pdf}
\end{minipage}%
}%
\subfigure[MoCoV2-Watermarked]{
\begin{minipage}[t]{0.25\linewidth}
\centering
\includegraphics[width=1.8in]{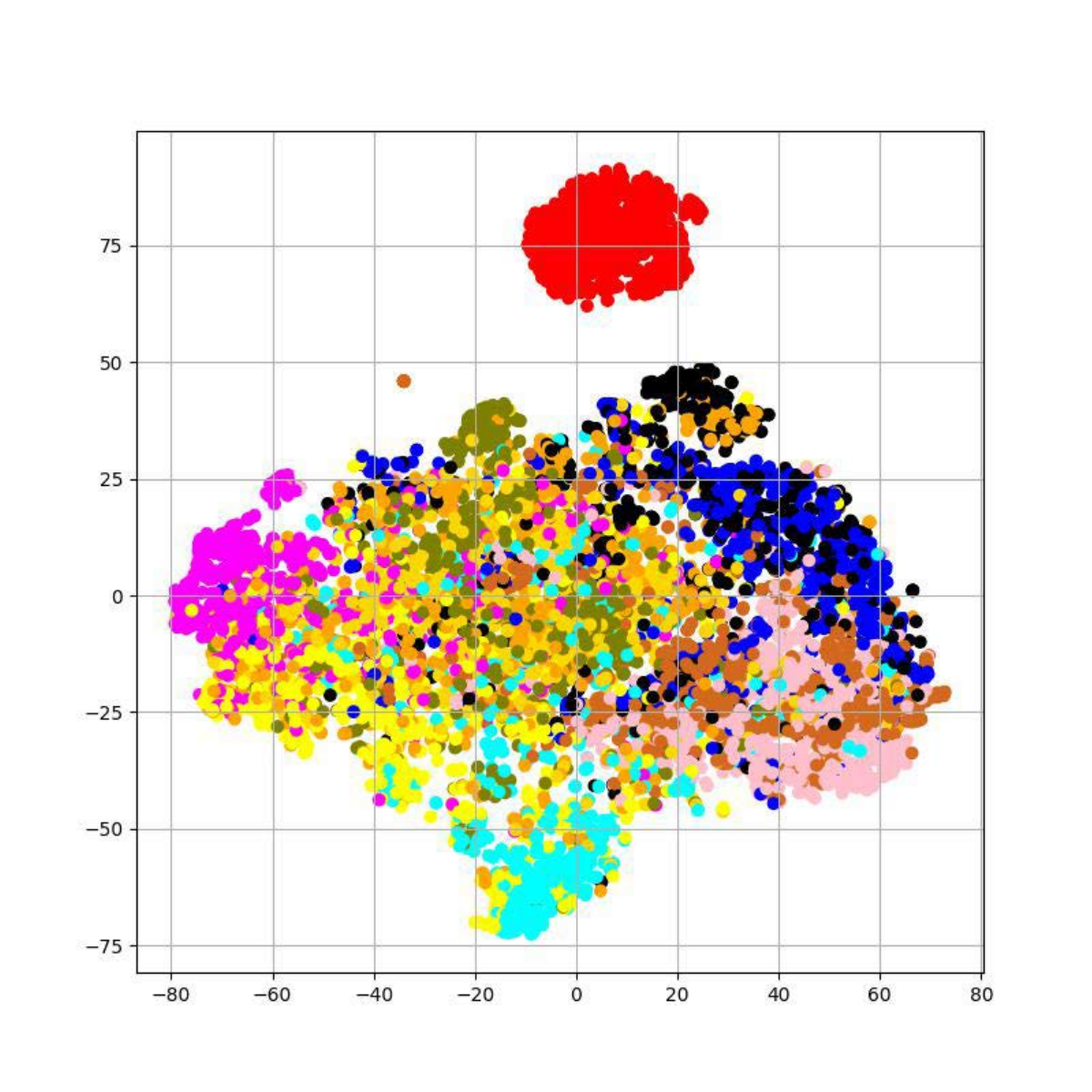}
\end{minipage}%
}%
\subfigure[BYOL-Watermarked]{
\begin{minipage}[t]{0.25\linewidth}
\centering
\includegraphics[width=1.8in]{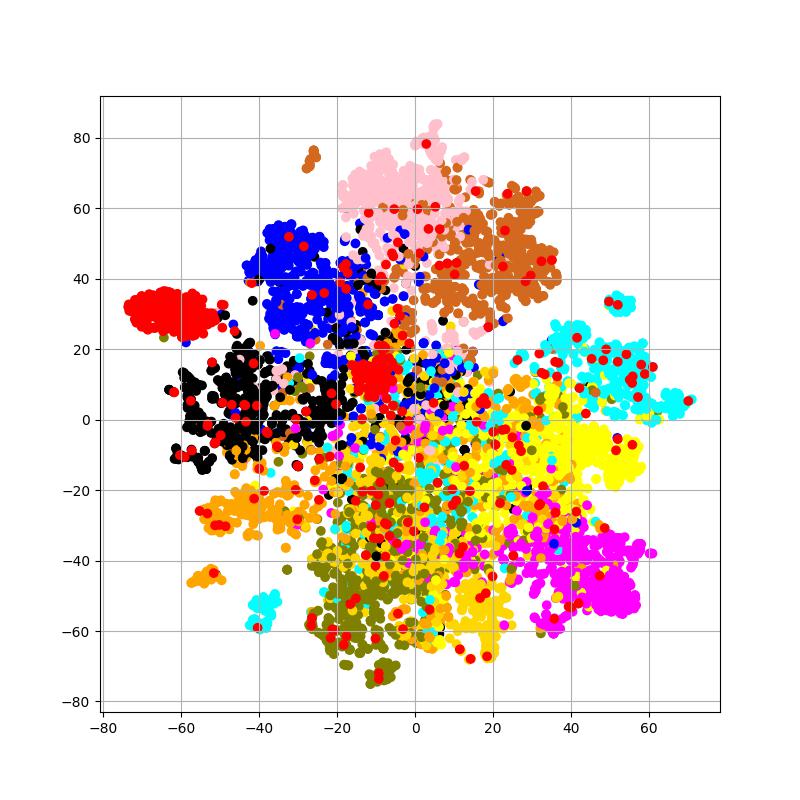}
\end{minipage}%
}%
\subfigure[BiGAN-Watermarked]{
\begin{minipage}[t]{0.25\linewidth}
\centering
\includegraphics[width=1.8in]{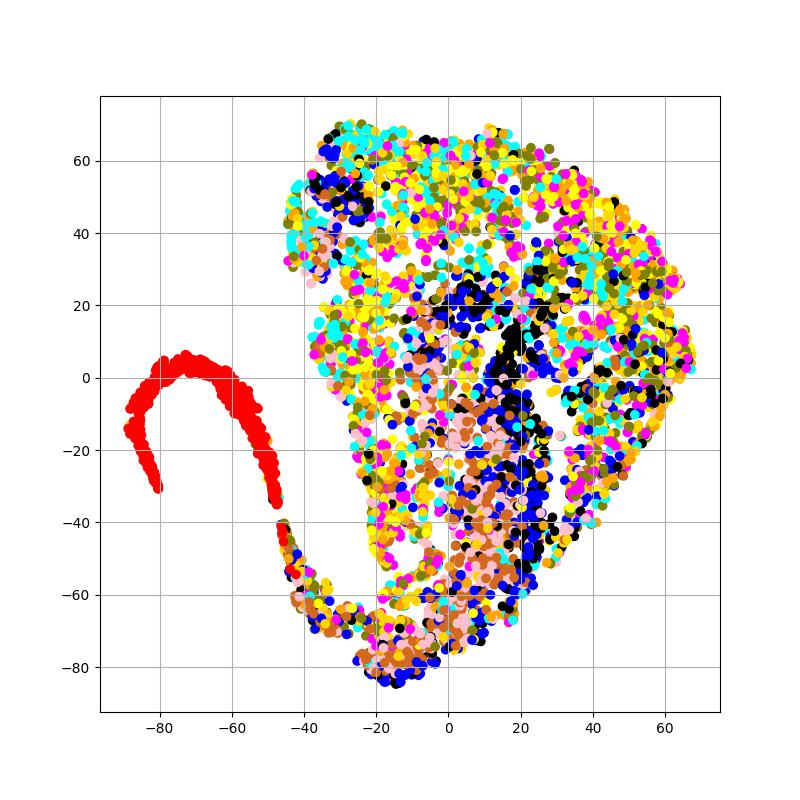}
\end{minipage}%
}%

\centering
\caption{Visualization of embedding space of watermarked/clean self-supervised models using T-SNE~\cite{hinton2002stochastic}. 
}
\label{fig:watermarked-four-ssl-models}
\end{figure*}

\end{document}